\documentclass[prb,aps,amssymb,amsmath,twocolumn]{revtex4-2}

\newcommand{\cotanh}{\ensuremath{\mathrm{cotanh}}}
\usepackage{graphicx} 
\usepackage{dcolumn}
\usepackage{hyperref}
\begin{document}
\title{One dimensional Bose-Hubbard model with long range hopping} 
\author{Edmond Orignac}
\affiliation{ENSL, CNRS, Laboratoire de Physique, F-69342 Lyon, France}
\date{\today}
\begin{abstract}
Interacting one-dimensional bosons with long range hopping decaying as
a power law $r^{-\alpha}$ with distance $r$ are considered with the renormalization group and the
self-consistent harmonic approximation. For $\alpha\ge 3$, the ground
state is always a Tomonaga-Luttinger liquid, whereas for $\alpha <3$,
a ground state with long range order breaking the continuous global gauge
symmetry becomes possible for sufficiently weak repulsion. At positive
temperature, continuous symmetry breaking becomes restricted to
$\alpha<2$, and for $2<\alpha<3$, a Tomonaga-Luttinger liquid with the
Tomonaga-Luttinger exponent diverging at low temperature is found. 
\end{abstract} 
\maketitle

\section{Introduction}\label{sec:introduction}
During the last twenty years, ultracold atomic
gases\cite{AdvPhys07-Lewenstein,Bloch2008,chanda_2025} and trapped ions\cite{johanning_2009,foss-feig_2025} have furnished
increasingly sophisticated experimental platforms for the realization 
of many-body model systems. In particular, long range interactions and
hoppings can be engineered in trapped ion systems\cite{feng2023} and
Rydberg atoms\cite{emperauger2025}. It has also been proposed to
realize long range hopping and interactions using time periodic drives
applied to ultracold atoms\cite{giergiel_2025}. 
In a more traditional condensed matter setting, there are
proposals\cite{yang2024} to realize antiferromagnetic
spin chains having
long range interactions with atomic chains of transition metal
atoms\cite{tung2011} such as Cr. Long range hopping and long range
interactions can lead to unusual phase transitions. Such effects can be
particularly marked in one dimension. 
Indeed, in one dimension, with short range hopping and interactions, a
continuous symmetry breaking (CSB) long range order (such as superfluidity
or N\'eel ordering) is not possible\cite{landau-statmech-english} even
in the ground
state\cite{hohenberg67_theorem,mermin_wagner_theorem,mermin_theorem,coleman1973,pitaevskii1991,gelfert2001}
unless the order parameter commutes with the Hamiltonian. Instead,
the ground state may exhibit quasi-long range order with correlation
functions decaying as a power law with
distance\cite{giamarchi_book_1d}, the so-called Tomonaga-Luttinger
liquid (TLL) 
state\cite{tomonaga50_1D_electron_gas,luttinger_model,haldane_effective_harmonic_fluid_approach}
or short range order\cite{lecheminant_revue_1d}. However,
with long range hopping or interaction decaying with distance as a
power law $\sim 1/r^\alpha$ the effective dimensionality is
altered, allowing long range
order\cite{dyson1969,thouless1969,fisher1972,defenu2023}.  Knowing how $\alpha$ affects the
stability of the CSB and the TLL correlations is thus a necessity to
interpret experiments\cite{feng2023,emperauger2025}. From the theory
side, in quantum Heisenberg ferromagnets, it was
found\cite{nakano1995} using Green's functions methods that long range
order would obtain at $T>0$ in $d$ dimensions when
$d\le \alpha \le 2d$. It has been rigorously
established\cite{bruno2001} later that in one and two dimensions, for
$\alpha\ge 2d$, long range order disappears at $T>0$. In the ground
state\cite{parreira97_longrange1d_neel}, and in one dimension with
$\mathrm{SU(2)}$ symmetry, Néel order was proved unstable for
$\alpha>3$. Also in one dimension, \cite{laflorencie2005} antiferromagnetic
spin-1/2 chains with non-frustrating long-range exchange and full
$\mathrm{SU(2)}$ symmetry were shown by Quantum Monte Carlo methods to
exhibit long range ordering for any $\alpha <2$ in the ground
state. In the ordered phase, a selfconsistent harmonic
approximation\cite{pires_easy_1995}, 
spin wave\cite{yusuf2004,laflorencie2005}
and large$-N$ calculations\cite{laflorencie2005} predicted a
dispersion $\omega(k) \sim k^{(\alpha-1)/2}$ for the Goldstone
bosons. In the case of anisotropic XXZ spin-1/2
chains,\cite{maghrebi2017}, renormalization group
calculations showed that long range order occurs at a critical
$\alpha_c<3$ that varies with exchange anisotropy. A phase diagram was
obtained from Density Matrix Renormalization group
calculations\cite{ren2020,schneider2022}, showing the variation of the
critical $\alpha_c$ with the exchange anisotropy. When the
interactions are frustrated, long range ordering is disfavored. For
instance, in the exactly solved Haldane-Shastry
chain\cite{haldane_inv_square,shastry_inv_square} the ground state
remains in the Tomonaga-Luttinger liquid phase. With long-range
interactions\cite{calogero69_model2,sutherland71_model1,schulz_wigner_1d,inoue_conformal_2006,casula06_coulomb1d_qmc}
decaying as $1/r^\beta$, the effects are less dramatic. For $\beta>1$,
the TLL is preserved\cite{inoue_conformal_2006,calogero69_model2,sutherland71_model1}, and for $\beta=1$, density-wave correlations
still present a quasi-long range order, albeit with a decay that is
slower than any power law. 
In the present manuscript, I 
consider the competition between CSB and TLL in a model of interacting one-dimensional bosons
with long range hopping. In Sec.~\ref{sec:model}, I will describe the model
and discuss its relation with antiferromagnetic XXZ spin-1/2
chains with unfrustrated long range exchange interactions.
In Sec.~\ref{sec:bosonization}, I will review the bosonized
representation of the model. In Sec.~\ref{sec:rg}, I will discuss its renormalization group
treatment\cite{maghrebi2017}. In Sec.~\ref{sec:scha}, I will apply the selfconsistent
harmonic approximation (SCHA)\cite{coleman_equivalence,suzumura_sg} to
describe the phase with continuous symmetry breaking both for $T=0$
and $T>0$. I will also discuss the calculation of the Luttinger
exponent and the case of frustrated hopping.

\section{Model and Hamiltonian}\label{sec:model}
Let's consider a Bose-Hubbard
model\cite{lewenstein07_coldatoms_review} on an infinite
one-dimensional lattice
\begin{equation}
  \label{eq:bose-hubbard}
  H_0=\sum_{j=-\infty}^{+\infty} \left[-J  (b^\dagger_j b_{j+1} + b^\dagger_{j+1} b_j) + \frac U 2 n_j (n_j-1)\right],  
\end{equation}
where $J>0$ is the transfer integral, $U>0$ is the on-site repulsion,
$b_j$ is the boson annihilation operator on site $j$, $n_j=b^\dagger_j
b_j$ the number operator.  Away from integer fillings, the Bose-Hubbard
model (\ref{eq:bose-hubbard}) presents a TLL ground
state\cite{monien98_bose_1d,kuhner_bose_hubbard_critical_point}.
The Bose-Hubbard model (\ref{eq:bose-hubbard}) is perturbed by a long range hopping decaying as a power law
\begin{equation}
  \label{eq:longrange_hopping}
  H_{LR} =-J_{LR} \sum_j \sum_{l=2}^{+\infty} \frac{b^\dagger_j b_{j+l} + b^\dagger_{j+l} b_j}{l^\alpha},  
\end{equation}
giving a total Hamiltonian $H=H_0+H_{LR}$. To preserve the
extensivity of the ground state energy\cite{defenu2023}, we have  to
restrict ourselves to $\alpha>1$. The long-range hopping favors the formation of
a superfluid phase in which $\langle b_j \langle \ne 0$, breaking the continuous
U(1) global gauge symmetry $b_j \to e^{i\gamma} b_j$. Such CSB/TLL
competition has a magnetic analogue.   
Indeed, in the hard core boson limit $U/J \to +\infty$,
the Bose-Hubbard model maps onto the ferromagnetic XY spin
chain\cite{matsubara1956,fisher_xxz} with $S_j^+=b_j^\dagger$,
$S_j^{-}=b_j$ and $S_j^z=b^\dagger_j b_j-1/2$. The long range
hopping~(\ref{eq:longrange_hopping}) becomes a ferromagnetic
long-range easy-plane exchange interaction. By a $\pi$ rotation around
the $z$ axis on odd sites, $S^{\pm}_j \to (-)^j S^{\pm}_j$ the nearest
neighbor exchange interaction becomes antiferromagnetic. 
Adding a nearest-neighbor Ising exchange,
\begin{equation}
  J_z \sum_j S_j^z S_{j+1}^z 
\end{equation}
turns the XY spin chain into the XXZ spin chain which is integrable by
Bethe Ansatz techniques\cite{orbach1959}. At zero magnetization, with
$|J_z|<|J|$, or with partial magnetization, its ground state is a
TLL\cite{luther_chaine_xxz,haldane_xxzchain}.
In the XY or XXZ spin chain, the long
range interaction~(\ref{eq:longrange_hopping}) takes the unfrustrated
form\cite{laflorencie2005}
\begin{equation}
  H_{LF}^{AF} = J_{LR} \sum_j \sum_{l=2}^{+\infty}
  \frac{(-1)^l}{l^\alpha} (S^+_j S^{-}_{j+l}+S^-_j S^{+}_{j+l}),   
\end{equation}
and the TLL phase is in competition with a Néel state with the spins
lying in the XY plane. The role of the global U(1) symmetry is now
played by O(2) rotations around the $z$ axis. Because of the
equivalence between the two models, in the following, we will
concentrate on the boson model
(\ref{eq:bose-hubbard})--~(\ref{eq:longrange_hopping}).

\section{Bosonization}\label{sec:bosonization}
In the absence of long-range hopping, and away from commensurate
filling, the Bose-Hubbard
model\cite{monien98_bose_1d,kuhner_bose_hubbard_critical_point,kuhner_bosehubbard,ejima2011}
has a Tomonaga-Luttinger liquid\cite{giamarchi_book_1d} ground
state. Its low-energy excitations and its correlation functions at low
energy, long wavelength are described by
bosonization\cite{haldane_bosons,efetov_larkin75,giamarchi_book_1d}.
The bosonized Hamiltonian away from integer filling
reads\cite{haldane_bosons,efetov_larkin75,giamarchi_book_1d}
\begin{equation}
  \label{eq:bosonized-bh}
  \mathcal{H}_0 = \int \frac{dx}{2\pi} \left[ u K (\pi \Pi)^2 + \frac{u}{K} (\partial_x \phi)^2\right],  
\end{equation}
where $[\phi(x),\Pi(y)]=i\delta(x-y)$, $u$ is the velocity of
excitations, and $K$ the Tomonaga-Luttinger exponent. In the hard core
boson limit, $K=1$ and $u=2Ja \sin(\pi n)$ for a filling of $n$ bosons
per site. The Tomonaga-Luttinger exponent has been calculated
numerically in the general case\cite{kuhner1998,kiely2022}, and it was
found that $K\ge 1$, with $K\to +\infty$ as $U\to 0$.  The bosonized
representations for the
operators\cite{haldane_bosons,giamarchi_book_1d} are
\begin{eqnarray}
  \label{eq:bosonized-bosons}
  b_j &=& e^{i \theta(j a)} \sum_{m=-\infty}^{+\infty} A_m e^{2i m (\phi(ja) - 2\pi n_0 j a)}, \\
  \label{eq:bosonized-density}
  \frac{b^\dagger_j b_j} a &=& n_0 -\frac 1 {\pi} \partial_x \phi(ja)
                               + \sum_{m=1}^{+\infty} B_m \cos [2m
                               (\phi (ja) - 2\pi n_0 ja)], \nonumber
  \\ 
\end{eqnarray}
where $A_m$ and $B_m$ are non-universal parameters that depend on microscopic details of the model, $a$ is the lattice spacing, $n_0 = \langle b^\dagger_j b_j\rangle/a$ is the number of bosons per unit length, and
\begin{equation}
  \label{eq:theta-def}
  \theta(x)=\pi \int_{-\infty}^x dy \Pi(y). 
\end{equation}
In the hard core boson limit,\cite{ovchinnikov2004_ff_xx} the
coefficient $A_0^2 =0.29417\ldots$. 
In the case of the XXZ spin chain, the same low-energy description
applies\cite{luther_chaine_xxz,haldane_xxzchain} and the coefficients
in Eqs.~(\ref{eq:bosonized-bosons})--~(\ref{eq:bosonized-density}) are
known analytically for zero
magnetization\cite{lukyanov_xxz_asymptotics,ovchinnikov2004_ff_xx} and numerically in
the partially magnetized
state\cite{furusaki_correlations_xxz_magneticfield}. 
The long range hopping, Eq.~(\ref{eq:longrange_hopping}), can be
expressed perturbatively using bosonization. Retaining only the most
relevant terms, we find 
\begin{eqnarray}
  \label{eq:lr-bosonized}
  \mathcal{H}_{LR} = -\frac{2 J_{LR} A_0^2}{a} \sum_{l=2}^{+\infty} \int dx \frac{\cos [\theta(x+la) -\theta(x)]}{l^\alpha}.  
\end{eqnarray}
The less relevant terms are of the form
\begin{eqnarray}
&&  -\sum_{m\ge 1} \sum_l \int dx \cos [\theta(x+la) -\theta(x)]
   \nonumber \\
  && \times \frac{\cos 2m [\phi(x+la) -\phi(x) -2\pi n_0 la]}{l^\alpha}.  
\end{eqnarray}
They are less relevant for two reasons. First, their scaling dimension
for the renormalization group is $\frac 1 {2K} + 2 m^2 K\ge 2m$, so
they will be at best marginal for $m=1$ and irrelevant for $m>1$.
Second, the factors
$e^{\pm 2i \pi m n_0 l a} $ make the sums
\begin{eqnarray}
  \sum_{l=2}^{+\infty} \frac{e^{\pm 2i \pi m n_0 l a}}{l^\alpha},  
\end{eqnarray}
convergent\cite{olver2010nist} for any $\alpha>0$, reducing their
contribution to an effective short range hopping at incommensurate
fillings.
To  understand the effect of the
most relevant terms, it is convenient to apply first the duality
transformation
\begin{eqnarray}u
  \label{eq:duality}
  \pi P = \partial_x \phi, \nonumber \\
  \lbrack\theta(x),P(y)\rbrack=i \delta(x-y) 
\end{eqnarray}
to rewrite
\begin{equation}
  \label{eq:bosonized-bh-dual}
  \mathcal{H}_0 = \int \frac{dx}{2\pi} \left[ \frac{u}{K} (\pi P)^2 + u K (\partial_x \theta)^2 \right],  
\end{equation}
and take the classical limit in the interaction term~(\ref{eq:lr-bosonized}).
In that limit, a Taylor expansion of the cosines gives 
\begin{eqnarray}
  \label{eq:lr-bosonized-taylor}
  \mathcal{H}_{LR} = -\frac{2 J_{LR} A_0^2}{a} \sum_{l=2}^{+\infty}
  \int dx \frac{1- [\theta(x+la) -\theta(x)]^2/2 +\ldots
  }{l^\alpha}. \nonumber \\   
\end{eqnarray}
and the contribution to the ground state energy is finite
only\cite{defenu2023} for $\alpha>1$.
Applying a second Taylor expansion\cite{maghrebi2017} to the terms $(\theta(x+l
a)-\theta(x))^2/l^a$, we find  
\begin{equation}
  \sum_l \frac{l^2 a^2}{l^\alpha} (\partial_x \theta)^2,  
\end{equation}
and for $\alpha>3$ this sum is convergent. Its contribution can then be
absorbed in a redefinition of $uK$. This suggests that the
Tomonaga-Luttinger liquid should be stable in that limit. For
$\alpha <3$, we have a divergent sum, so the second Taylor expansion
does not make sense. This hints that the long range hopping
can destabilize the Tomonaga-Luttinger liquid when $\alpha<3$.
Of course, since we are dealing with operators
instead of classical variables, we cannot actually apply a
straightforward Taylor expansion to the cosines in
Eq.~(\ref{eq:lr-bosonized}). To go beyond a heuristic reasoning,
we will use first a
renormalization group approach\cite{maghrebi2017} to consider the stability of the
Tomonaga-Luttinger liquid in Sec.~\ref{sec:rg}. Then, to describe the
case in which the Luttinger liquid fixed point is unstable, we will
turn  to  the SCHA in
Sec.~\ref{sec:scha}. 

\section{Renormalization group treatment}\label{sec:rg}
\begin{figure}[h]
  \centering
  \includegraphics[width=9cm]{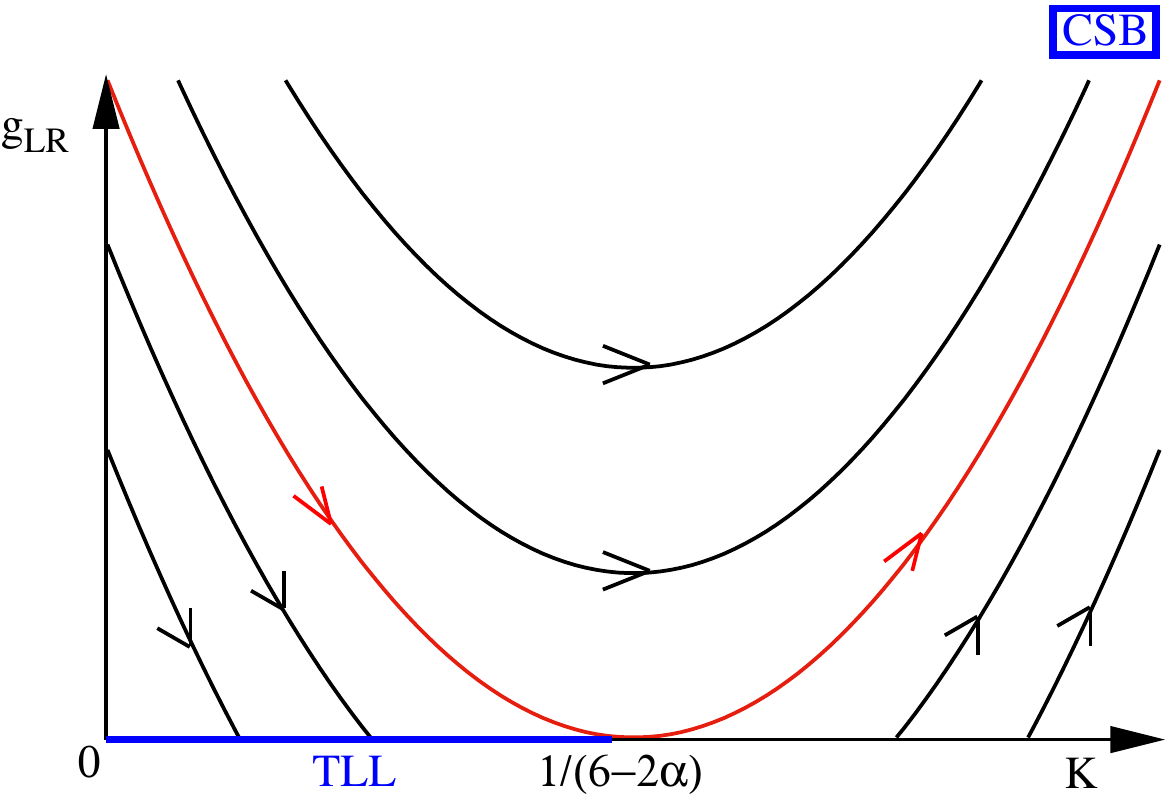}
  \caption{Renormalization group flow of
    Eqs.~(\ref{eq:rg-reduced}). The Tomonaga-Luttinger liquid (TLL) fixed line (blue)
    corresponds to $g_{LR}=0$ and $0<K<1/(6-2\alpha)$. The separatrix
    $g_{LR}=(3-\alpha)[K-1/(6-2\alpha)]^2/2$ is represented in
    red. When $K(0)<1/(6-2\alpha)$ and the initial point
    $(K(0),g_{LR}(0))$ is on the separatrix or below it, the flow
    terminates on the Tomonaga-Luttinger liquid (TLL) fixed line. In other cases, the
    flow goes to strong coupling, indicating the formation of the
    continuous symmetry beaking (CSB). }
  \label{fig:rgflow}
\end{figure}

In order to apply a renormalization group approach, we have first to replace
the discrete summation in Eq.~(\ref{eq:lr-bosonized}) by an
integration to  write  the
Matsubara action\cite{maghrebi2017,schneider2022,dupuis2024} with a
triple integral. When the characteristic lengthscale for the variation of $\theta$
is much larger than the lattice spacing, replacing the discrete
variable $la$ by a continuum variable $(x-y)$ is justified. This
will be valid in particular in the vicinity of the CSB to TLL phase
transition. The resulting action is
\begin{widetext} 
\begin{eqnarray}
  \label{eq:ls-boso-contin}
S=\int \frac{K dx d\tau}{2\pi}\left[ u(\partial_x \theta)^2 +
  \frac{(\partial_\tau \theta)^2} u \right]  -\frac{J_{LR} A_0^2}{a^{2-\alpha}}  \int d\tau \int_{|x-y|\ge a} dx dy \frac{\cos [\theta(x,\tau) -\theta(y,\tau)]}{|x-y|^\alpha}. 
\end{eqnarray}
\end{widetext}
For the renormalization group procedure, we use a real
space cutoff such that in any operator product $O_1(x_1,\tau_1)
O_2(x_2,\tau_2)$, we impose $(x_1-x_2)^2 + u^2 (\tau_1-\tau_2)^2 >
a^2$. Such scheme permits the use of the Operator Product
Expansion (OPE)\cite{cardy_book_renormalization} approach to derive
the renormalization group equations of $u$ and $K$.   
From the scaling dimension\cite{maghrebi2017,schneider2022,dupuis2024} of the operator
$\cos\theta$,
the flow equation for $J_{LR}$ is 
\begin{equation}
  \label{eq:jlr-rg}
  \frac{dJ_{LR}}{d\ell} = \left(3-\alpha-\frac 1 {2K}\right) J_{LR},  
\end{equation}
and $J_{LR}$ is relevant only if $\alpha+\frac{1}{2K} < 3$. In particular, when $\alpha\ge 3$, no CSB is
possible,in agreement with\cite{parreira97_longrange1d_neel}.     
When $J_{LR}$ is relevant, and $K$ is sufficiently larger than $1/(6-2\alpha)$, the flow can be treated as a vertical flow, leading to a characteristic lengthscale
\begin{equation}\label{eq:corrlength-rg}  
 \xi_{\mathrm{RG}} \sim a \left(\frac{\pi K J_{LR} A_0^2
     a}{u}\right)^{-\frac{1}{3-\alpha-\frac 1 {2K}}}
\end{equation}
beyond which the superfluid CSB order is established. The replacement
of the discrete sum by the continuum expression in the
action~(\ref{eq:ls-boso-contin}) is justified when $\xi_{\mathrm{RG}}
\gg a$.  As $K$ is getting
closer to $1/(6-2\alpha)$, the renormalization of $K$ during the flow
becomes too important to be neglected.\cite{maghrebi2017}
Using the OPE approach (see App.~\ref{app:ope-rg})
we obtain the renormalization group equations for $u$ and $K$ in the
form 
\begin{eqnarray}
  \label{eq:u-rg}
  \frac{du}{d\ell}&=&\frac{\pi J_{LR} A_0^2  a}{K},  \\
  \label{eq:K-rg} 
  \frac{dK}{d\ell}&=&\frac{\pi J_{LR} A_0^2 a}{u}.
\end{eqnarray}
Those equations differ from the ones in Ref.~\onlinecite{maghrebi2017}
by the constant multiplying $J_{LR}$. The reason is that we are using
a real-space cutoff, while in Ref.~\onlinecite{maghrebi2017} a
momentum-space cutoff was used. This simply means that the initial
values of $u,K$ and $J_{LR}$ for a given microscopic model will depend
also on the chosen regularization scheme.   
The ratio $u(\ell)/K(\ell)$ is invariant under the renormalization
group\cite{maghrebi2017}, so we can restrict the running variables to
$K$ and $J_{LR}$.
It is convenient to introduce the dimensionless coupling constant
\begin{equation}
  \label{eq:glr-def}
  g_{LR}=\frac{\pi K J_{LR} A_0^2 a}{u}, 
\end{equation}
and write the reduced RG equations in a form analogous to the Giamarchi-Schulz RG equations\cite{giamarchi_loc_lettre,giamarchi_loc}
\begin{eqnarray}
  \label{eq:rg-reduced}
  K \frac{dK}{d\ell} &=& g_{LR}, \nonumber \\
  \frac{d g_{LR}}{d\ell} &=& \left(3-\alpha -\frac{1}{2K}\right) g_{LR}.
\end{eqnarray}
Those equations~(\ref{eq:rg-reduced}) possess the invariant
\begin{eqnarray}\label{eq:rg-invariant} 
  \frac{\mathcal{C}}{2}=-\frac{3-\alpha} 2 K^2 +\frac K 2 + g_{LR},  
\end{eqnarray}
which allows to integrate the RG equation for $K(\ell)$ in the
form~(see App.~\ref{app:integration-rg}) 
\begin{eqnarray}\label{eq:rg-integrated} 
  \int_{K(0)}^{K(\ell)} \frac{2 K dK}{(3-\alpha) K^2 - K + \mathcal{C}}= \ell.   
\end{eqnarray}
Let's consider first the case of $\alpha<3$, where $g_{LR}$ can be
relevant. The behavior of the renormalization group flow is
represented on Fig.~\ref{fig:rgflow}. As in the Giamarchi-Schulz case, the flow lines are parabolas instead
of the hyperbolas of the
Kosterlitz-Thouless\cite{kosterlitz_thouless,kosterlitz_renormalisation_xy}
RG flow.  
When $\mathcal{C}=[4(3-\alpha)]^{-1}$, the flow is sitting of the
separatrix in Fig.~\ref{fig:rgflow}. if $K(0)<1/(6-2\alpha)$, it terminates in the TLL phase
 with a fixed point exponent $K^*=1/(6-2\alpha)$. The correlation function
\begin{equation}
  \langle e^{i\theta(x)} e^{-i\theta(0)} \rangle = \mathcal{A} \left(\frac{a}{|x|}\right)^{3-\alpha} (\ln |x|/a)^{-2},  
\end{equation}
where $\mathcal{A}$ is a constant, shows logarithmic corrections that
slightly reduce the quasi-long range order with respect to the fixed
point behavior. 
When $\mathcal{C}>[4(3-\alpha)]^{-1}$, the flow lies above the
separatrix on Fig.~\ref{fig:rgflow} and $K(\ell)$ and $g_{LR}$ always
flow to $+\infty$. This corresponds to the CSB
phase\cite{maghrebi2017,schneider2022,dupuis2024}. When
$\mathcal{C}<[4(3-\alpha)]^{-1}$, the flow lies below the separatrix
on Fig.~\ref{fig:rgflow}.  When $K(0)>1/(6-2\alpha)$ both
$g_{LR}(\ell)$ and $K(\ell)$ diverge giving again a CSB phase. When
$K(0)<1/(6-2\alpha)$, the flow terminates at $g_{LR}^*=0$ and
\begin{equation}\label{eq:K-fixedpoint} 
  K^*=\frac{1}{2(3-\alpha)}
  -\sqrt{\left(\frac{1}{2(3-\alpha)}-K(0)\right)^2
    -\frac{2g_{LR}(0)}{3-\alpha}},  
\end{equation}
 yielding a  TLL phase. In
  particular, if we start at $K=1/(6-2\alpha)$, the smallest $J_{LR}$
  perturbation induces superfluid order, as  was observed
  in the $\mathrm{SU(2)}$ invariant case with $\alpha=2$ \cite{laflorencie2005}. 
 For fixed $J_{LR}$, Eq.~(\ref{eq:rg-integrated}) predicts a critical value of $K$,
\begin{eqnarray}\label{eq:rg-kcrit} 
  K_c &=&\frac{1}{2(3-\alpha)} \left[1 - \sqrt{\frac{4 \pi J_{LR}
        A_0^2a}{u} +\left(\frac{2 \pi J_{LR}  A_0^2a}{u}
      \right)^2}\right. \nonumber \\
  && \left. +  \frac{2 \pi J_{LR}  A_0^2
        a}{u} \right] 
\end{eqnarray}
such that the CSB/TLL phase transition takes place at $K=K_c$. For
fixed K $K$, Eq.~(\ref{eq:rg-integrated}) gives the CSB/TLL phase
transition at $g_{LR}^c=(3-\alpha)(K-1/(6-2\alpha))^2/2$.
For $g_{LR}\agt g_{LR}^c$, the characteristic lengthscale behaves as
\begin{equation}
  \xi_{\mathrm{RG}} \sim a \exp\left[\frac{\mathcal{A}}{\sqrt{g_{LR}-g_{LR}^c}}\right],  
  \end{equation}
  as in a Kosterlitz-Thouless\cite{kosterlitz_thouless} transition.
Using
Eq.~(\ref{eq:glr-def}), we find in the case of hard core bosons at
half-filling, with $u=2J a$ and $K=1$ that for a fixed $J_{LR}$, the
phase transition takes place at $\alpha=\alpha_c>5/2$ with 
\begin{equation}
   \frac{(2\alpha_c -5)^2}{3-\alpha_c}
  =2 \pi\frac{J_{LR}}{J} \times (0.29417\ldots)^2.  
\end{equation}
In the case of a Heisenberg antiferromagnetic spin-$1/2$ chain, with
zero magnetization\cite{orignac04_spingap}, $u=\pi J a/2$ and $K=1/2$ with $A_0^2 \simeq
(2\pi^3)^{-1/4}$ we have $2<\alpha_c<3$ and 
\begin{equation}
   \frac{(2-\alpha_c)^2}{3-\alpha_c}
  =8 \frac{J_{LR}}{(2\pi^3)^{1/2} J}.
\end{equation}
In general, we find $\alpha_c-3+1/(2K) \sim (J_{LR}/J)^{1/2}$, so
$\alpha_c \simeq 3-1/(2K)$ requires to make $J_{LR}$ very small
compared with the nearest neighbor hoping $J$.    

For $\alpha=3$, $g_{LR}$ is irrelevant, and the invariant of
Eq.~(\ref{eq:rg-invariant}) reduces to $C=K(0)+2g_{LR}(0)$. Integrating the
renormalization group equations leads to
\begin{eqnarray} 
K(\ell)&=&K(0)+2g_{LR}(0)(1-e^{-\frac{\ell+4g_{LR}(0)}{2K(0)+4g_{LR}(0)}})  \\
&& +o(e^{-\ell/(2K(0)+4g_{LR}(0))}), 
\end{eqnarray}
 indicating that the TLL fixed point with $K^*=K(0)+2g_{LR}(0)$ is reached
quickly, without any logarithmic correction. This is in agreement with the experimental results with dipolar interactions in Rydberg atoms\cite{emperauger2025}.
For $\alpha>3$, we have $C>0$ and in Eq.~(\ref{eq:rg-integrated}), the
denominator has to remain positive. For $\ell \to +\infty$, $K(\ell)$
goes to the only positive zero of the denominator in Eq.~(\ref{eq:rg-integrated}), so the fixed point value is
\begin{equation}
  K^*=2 \frac{(\alpha-3) K(0)^2 +K(0)
    +2g_{LR}(0)}{\sqrt{1+4(\alpha-3)[(\alpha-3) K(0)^2 +K(0)
      +2g_{LR}(0)]} +1}.   
\end{equation}

\section{Self-consistent harmonic approximation}\label{sec:scha}
In Sec.~\ref{sec:bosonization}, we have seen that the TLL phase could
become unstable in the presence of the long range
hopping~(\ref{eq:longrange_hopping}). 
In the present section, we apply the self-consistent Harmonic
Approximation\cite{coleman_equivalence,suzumura_sg,watanabe_2ch,donohue_thesis,citro04_spinpeierls,cazalilla_deconfinement_long,you2012,foini_ladder_quench_cold,majumdar2023}
to the long range hopping~(\ref{eq:lr-bosonized}) in order to
calculate the TLL exponent or describe
the CSB phase. We emphasize that the self-consistent harmonic
approximation used in the present article is different from the one
used in Ref.~\onlinecite{pires_easy_1995}. In
Ref.~\onlinecite{pires_easy_1995}, a Villain semi-polar representation of the
spin operators\cite{villain1974} is used, and the nearest neighbor
exchange interactions are approximated. By contrast, we are treating
exactly the nearest neighbor interaction, and treating only the
interactions beyond nearest neighbor within the approximation.
In the self-consistent Harmonic approximation (SCHA), we rewrite the operator
\begin{eqnarray}
  \label{eq:normal-order}
  \cos [\theta(x+la)-\theta(x)]&=&\langle \cos
  [\theta(x+la)-\theta(x)]  \rangle \nonumber \\  
  && \times : \cos [\theta(x+la)-\theta(x)]:,  
\end{eqnarray}
where $\langle \ldots \rangle $ stands for the expectation value, and $:\ldots:$ stands for normal ordering\cite{abrikosov_book}. The normal ordered expression has a Taylor expansion
\begin{equation}
  \label{eq:normal-Taylor}
  :\cos [\theta(x+la)-\theta(x)]: = \sum_{n=0}^{+\infty} \frac{(-)^n}{(2n)!} :(\theta(x+la)-\theta(x))^{2n}:, 
\end{equation}
and in the SCHA that expression is truncated at second order to obtain a quadratic Hamiltonian\cite{suzumura_sg}. The average in Eq.~(\ref{eq:normal-order}) is then calculated for that quadratic Hamiltonian yielding a self-consistent equation\cite{suzumura_sg}.
The SCHA Hamiltonian thus takes the form
\begin{eqnarray}
  \label{eq:scha-hamiltonian}
  \mathcal{H}_{SCHA}&=& \int \frac{dx}{2\pi} \left[ \frac{u}{K} (\pi P)^2 + u K (\partial_x \theta)^2 \right]  \\
  && + \frac{J_{LR}A_0^2} a \int dx \sum_{l=2}^{+\infty} \frac{g(l)}{l^\alpha} (\theta(x+la)-\theta(x))^2,  \nonumber
\end{eqnarray}
with
\begin{eqnarray}
  \label{eq:scha-condition}
  g(l)=\exp\left[-\frac{1}{2} \langle (\theta(x+la)-\theta(x))^2 \rangle_{H_{SCHA}} \right].  
\end{eqnarray}
In contrast with the renormalization group approach of
Sec.~\ref{sec:rg}, we don't need to turn the sum into an integral. For
that reason, we can expect that the SCHA gives more accurate results than the
RG when the characteristic length is not very large compared with the
lattice spacing. We will first consider  the ground state, discussing the
TLL and CSB
phases, then we will turn to the effect of temperature in Sec.~\ref{sec:temperature}.   
To diagonalize the quadratic Hamiltonian~(\ref{eq:scha-hamiltonian}),
we introduce the Fourier decomposition
\begin{eqnarray}
  \label{eq:fourier}
  \theta(x) &=& \frac{1}{\sqrt{L}} \sum_k e^{i k x} \theta(k),\nonumber \\
  P(x) &=&  \frac{1}{\sqrt{L}} \sum_k e^{i k x} P(k),
\end{eqnarray}
with $[\theta(k),P(-k')]=i\delta_{kk'}$ to rewrite
\begin{widetext} 
\begin{eqnarray}
  \label{eq:scha-fourier}
   \mathcal{H}_{SCHA}&=&\sum_k \left[\frac {\pi u}{2K} P(k) P(-k) +
                         \left(\frac{uK k^2}{2\pi} + \frac{J_{LR} A_0^2}{a} \sum_{l=2}^{+\infty} \frac{g(l)}{l^\alpha} |1-e^{ikla}|^2 \right) \theta(k) \theta(-k) \right].   
\end{eqnarray}
\end{widetext}
Now, define
\begin{eqnarray}
  \label{eq:dispersion}
  \omega(k)^2=u^2 k^2 +\frac{2\pi J_{LR} A_0^2 u}{Ka} \sum_{l=2}^{+\infty} \frac{g(l)}{l^\alpha} |1-e^{ikla}|^2,   
\end{eqnarray}
and rewrite
\begin{eqnarray}
  \mathcal{H}_{SCHA}&=&\frac 1 2 \sum_k \omega(k) \left[\frac{\pi
                        u}{K\omega(k)} P(k) P(-k) \right. \nonumber \\
  && \left. +\frac{K\omega(k)}{\pi u} \theta(k) \theta(-k) \right]. 
\end{eqnarray}
With the help of the bosonic creation and and annihilation operators,
\begin{eqnarray}
  \label{eq:scha-oscillator}
  a(k)&=&\left(\frac{K\omega(k)}{2\pi u}\right)^{1/2} \theta(k) + i
          \left(\frac{\pi u}{2 K\omega(k)}\right)^{1/2} P(k), \\
  a^\dagger(k)&=&\left(\frac{K\omega(k)}{2\pi u}\right)^{1/2} \theta(k) - i \left(\frac{\pi u}{2K\omega(k)}\right)^{1/2} P(k). 
\end{eqnarray}
we obtain the final form of the SCHA Hamiltonian,
\begin{eqnarray}
   \mathcal{H}_{SCHA}= \sum_k \omega(k) \left( a^\dagger(k)a(k) +\frac 1 2 \right).  
\end{eqnarray}
Inverting Eq.~(\ref{eq:scha-oscillator}) yields
\begin{eqnarray}
  \label{eq:scha-fields} 
  \theta(k)&=&\left(\frac{\pi u}{K\omega(k)}\right)^{1/2} \frac{a(k)+a^\dagger(-k)}{\sqrt{2}}, \nonumber \\
  P(k)&=& \left(\frac{K\omega(k)}{\pi u}\right)^{1/2} \frac{a(k)-a^\dagger(-k)}{i\sqrt{2}} 
\end{eqnarray}
and in the ground state,  Eq.~(\ref{eq:scha-condition}) gives
\begin{eqnarray}
  \label{eq:scha-g-l}
  g(l)=\exp\left[-\int_0^{+\infty} \frac{d(uk)}{2 K \omega(k)} (1-\cos (kla)) e^{-ka} \right].
\end{eqnarray}
In the integral~(\ref{eq:scha-g-l}), we have introduced the factor $e^{-ka}$ to take into account the cutoff on momentum at $\sim \frac{\pi}a$ resulting from the presence of a lattice.
The equations~(\ref{eq:dispersion}) and (\ref{eq:scha-g-l}) form a selfconsistent set of equations determining the sequence $g(l)$ via $\omega(k)$.
We now have to analyze the behavior of $\omega(k)$ as $k\to 0$ for a given sequence $g(l)$.

\subsection{Tomonaga-Luttinger liquid phase}\label{sec:TLL}

When the series
\begin{eqnarray}
 \sum_{l=2}^{+\infty} \frac{g(l)}{l^{\alpha-2}}.  
\end{eqnarray}
is convergent, we have the limiting behavior as $k\to 0$, 
\begin{eqnarray}
   \sum_{l=2}^{+\infty} \frac{g(l)}{l^\alpha}(1-\cos (kla)) =
  \frac{k^2 a^2}{2} \sum_{l=2}^{+\infty} \frac{g(l)}{l^{\alpha-2}} +
  o(k^2),  
\end{eqnarray}
implying $\omega(k)=\tilde{u}|k|$. 
The SCHA Hamiltonian~(\ref{eq:scha-fourier}) is then a Tomonaga-Luttinger liquid Hamiltonian with Tomonaga-Luttinger parameters $\tilde{u}$ and $\tilde{K}$ given by
\begin{eqnarray}\label{eq:uoverK} 
  \frac{\tilde{u}}{\tilde{K}} &=& \frac{u}{K} \\
  \label{eq:utimesK}
  \tilde{u}\tilde{K}&=& u K + 2\pi J_{LR} A_0^2 a \sum_{l=2}^{+\infty} \frac{g(l)}{l^{\alpha-2}} 
\end{eqnarray}
Porting that result in Eq.~(\ref{eq:scha-g-l}), yields $g(l) = l^{-1/(2\tilde{K})}$ leading to the self-consistent equation for $\tilde{K}$
\begin{eqnarray}\label{eq:scha-TLL} 
  \tilde{K} = K\sqrt{1+\frac{2\pi J_{LR} A_0^2 a}{uK}\left[\zeta\left(\alpha +\frac 1 {2 \tilde{K}}-2 \right)-1\right] },  
\end{eqnarray}
where $\zeta$ is the Riemann zeta
function\cite{olver2010nist}. Eq.~(\ref{eq:scha-TLL}) is defined
provided $3-\alpha -(2\tilde{K})^{-1}<0$. This condition is  similar
to the condition for irrelevance of long range hopping
Eq.~(\ref{eq:jlr-rg}). 
More precisely, according to Eq.~(\ref{eq:scha-TLL}),
$\tilde{K}>K$, $3-\alpha-(2K)^{-1}<3-\alpha -(2\tilde{K})^{-1}$ and the
irrelevance of $J_{LR}$ is a necessary condition for the existence of
a solution of Eq.~(\ref{eq:scha-TLL}).
If we expand Eq.~(\ref{eq:scha-TLL}) and Eq.~(\ref{eq:K-fixedpoint})
to first order in $J_{LR}/u$ and make $\alpha -3 +1/(2K) \ll 1$, the
two expressions coincide. This indicates that the continuum
action~(\ref{eq:ls-boso-contin}) and its renormalization treatment
give a good approximation near the CSB/TLL critical point.  
The geometric interpretation of
Eq.~(\ref{eq:scha-TLL}) is the intersection of the straight line $y=\tilde{K}$ with the curve
\begin{equation}\label{eq:curve-scha-tll} 
  y= K\sqrt{1+\frac{2\pi J_{LR} A_0^2 a}{uK}\left[\zeta\left(\alpha +\frac 1 {2 \tilde{K}}-2 \right)-1\right] }. 
\end{equation}
The right hand side of Eq.~(\ref{eq:curve-scha-tll}) goes to $K$ when
$\tilde{K}\to 0$, and increases with $\tilde{K}$.  For $\alpha>3$,
when $\tilde{K} \to +\infty$, it has the limit  $K\sqrt{1+2\pi J_{LR}
  A_0^2 a(\zeta(\alpha-2)-1)/(uK)}$ .  As a result, the straight line
$y=\tilde{K}$ always has a unique intersection with the curve defined
by Eq.~(\ref{eq:curve-scha-tll}), showing the stability of the TLL for
any $J_{LR}$, as found with the renormalization group.  
For $\alpha <3$, by contrast, the curve defined by
Eq.~(\ref{eq:curve-scha-tll}) exists only for
$\tilde{K}< 1/(6-2\alpha)$ and has a vertical asymptote when
$\tilde{K} \to 1/(6-2\alpha)$. In such case, either the curve does not
intersect the straight line at all, or it intersects it twice. In the latter
case, only the
lowest value of $\tilde{K}$ corresponds to a physical solution which  
exists only for $\tilde{K}<\tilde{K}_c$. The critical
value $\tilde{K}_c$ is obtained when the straight line $y=\tilde{K}$
reaches the limiting position where it is tangent to the
curve defined by Eq.~(\ref{eq:curve-scha-tll}). The
equality of the derivatives gives  
\begin{equation}
  \frac{dy}{d\tilde{K}}=-\frac{ \frac{\pi J_{LR} A_0^2 a}{2u \tilde{K}_c^2} \zeta'\left(\alpha +\frac{1}{2\tilde{K}_c}-2\right) }{\left[1+\frac{2\pi J_{LR} A_0^2 a}{uK}\left(\zeta\left(\alpha +\frac{1}{2\tilde{K}_c}-2\right)-1\right) \right]^{1/2}} = 1,   
\end{equation}
and combining that condition with Eq.~(\ref{eq:scha-TLL}), the relation
\begin{equation}
   \frac{\pi J_{LR} K A_0^2 a}{2u \tilde{K}_c^3} \zeta'\left(\alpha +\frac{1}{2\tilde{K}_c}-2\right)=-1, 
\end{equation}
determines $\tilde{K}_c<1/(6-2\alpha)$, which is the maximal value the TLL exponent
can take in the presence of long range hopping for given
$J_{LR},\alpha,u$ and $K$ within the SCHA.  
Since $J_{LR}a/u \ll 1$,  $\tilde{K_c}\alt 1/(6-2\alpha)$ one can
estimate $\tilde{K_c}$ by the approximation $\zeta'(s) \simeq
-1/(s-1)^2$. Such approximation leads to
\begin{equation}
  \tilde{K}_c \simeq \frac 1 2 \frac{1}{3-\alpha + \left(\frac{2\pi
        (3-\alpha)^4 J_{LR} A_0^2 a} u \right)^{1/2}+\ldots}, 
\end{equation}
and the maximal $\tilde{K}$ deviates from $1/(6-2\alpha)$ by an amount $O(J_{LR}u/a)^{1/2}$
So, in contrast with the 
renormalization group, which predicted that the Tomonaga-Luttinger
exponent could reach $1/(6-2\alpha)$ at the CSB/TLL phase transition, the
SCHA predicts a maximal exponent that is always strictly inferior to $1/(6-2\alpha)$. The
reason for such discrepancy is that in the SCHA the curvature of
the renormalization group flow near the critical point is neglected.   

To study the stability of the Luttinger liquid as a function of $K$, it is convenient to rewrite Eq.~(\ref{eq:scha-TLL}) in the form
\begin{widetext} 
\begin{equation}
  \label{eq:scha-tll-improved}
  K=\frac{\tilde{K}^2}{\sqrt{\tilde{K}^2+\left(\frac{\pi J_{LR} A_0^2 a}{u}\right)^2 \left[\zeta\left(\alpha +\frac 1 {2\tilde{K}} -2\right)-1\right]} + \frac{\pi J_{LR} A_0^2 a}{u} \left[\zeta\left(\alpha +\frac 1 {2\tilde{K}} -2\right)-1\right] }.  
\end{equation}
\end{widetext}
The right hand side reaches its maximum at $\tilde{K}=\tilde{K}_c$
which determines $K_m$ the maximum value of $K$ compatible with a
TLL phase. This leads to the estimate 
\begin{eqnarray}
K_m&\simeq& \frac{1}{2(3-\alpha)} \left[1-\left(\frac{2\pi J_{LR}A_0^2
            a}u\right)^{1/2} \left(3-\alpha +\frac 1 {3-\alpha}\right)
            +\ldots \right], \nonumber \\ 
\end{eqnarray}
which is similar to  Eq.~(\ref{eq:rg-kcrit}) for $J_{LR} \ll u/a$,
albeit with different prefactors for the term $O(J_{LR}
a/u)^{1/2}$. The prefactor given by SCHA is the largest, so the SCHA
is underestimating the critical value of $K$ compatible with the TLL
phase.    
\begin{figure}[h]
  \centering
  \includegraphics[width=9cm]{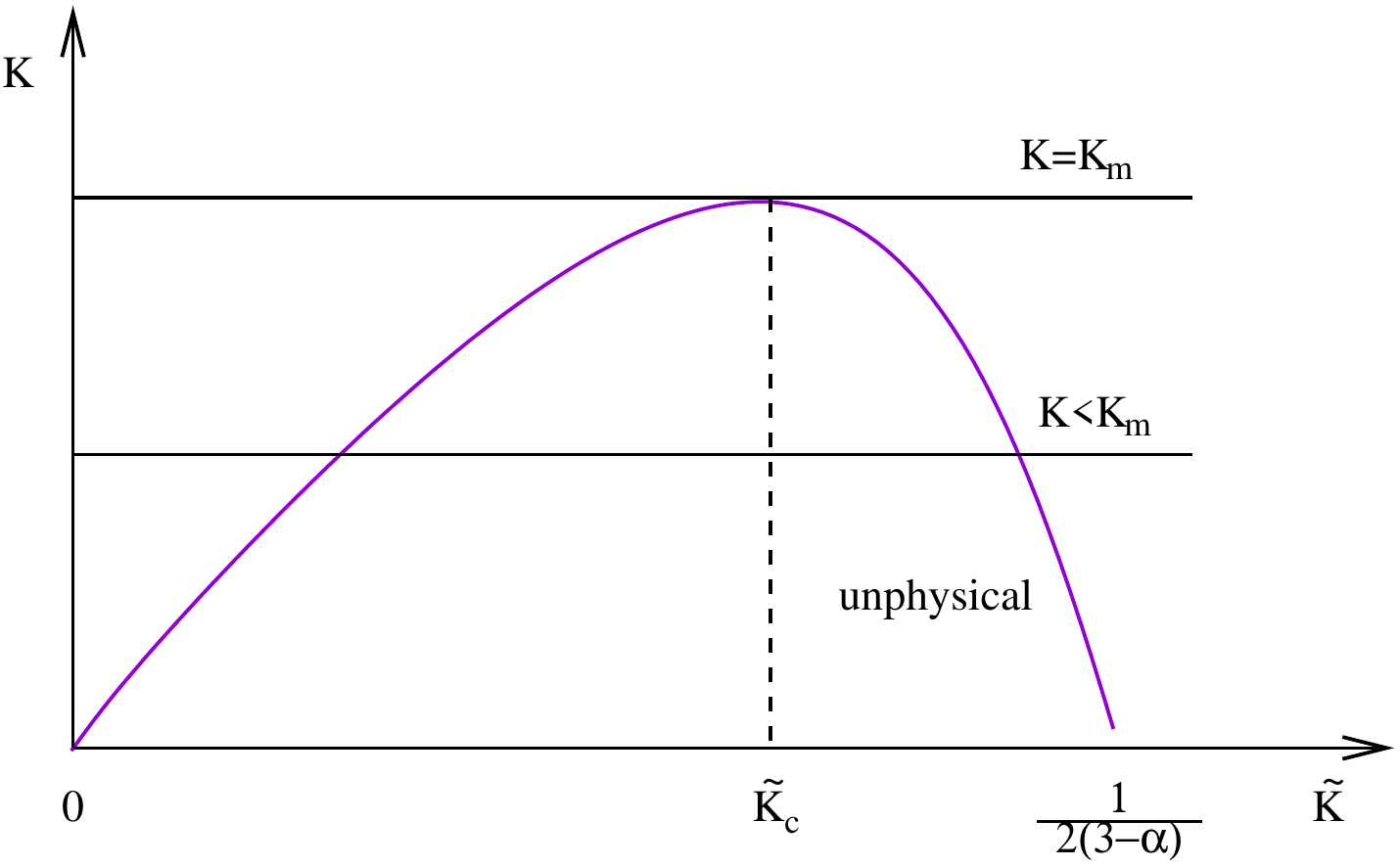} 
  \caption{Plot of Eq.~(\ref{eq:scha-tll-improved}) as a function of $\tilde{K}$ for $\alpha=2.8$ and $\pi J_{LR}A_0^2 a/u=0.1$. The expression is maximal at $\tilde{K}=\tilde{K}_c$ (indicated by a dashed vertical line). For $K>K_m$, Eq.~(\ref{eq:scha-tll-improved}) has no solution. For $K<K_m$, it has two solutions, the one with $\tilde{K}>\tilde{K}_c$ being unphysical. }
  \label{fig:Tllmax}
\end{figure}

\begin{figure}[h]
  \centering
  \includegraphics[width=9cm]{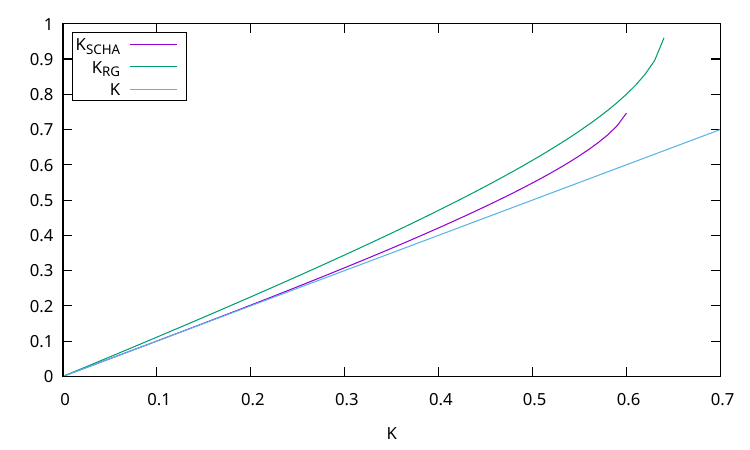} 
  \caption{Plot of the Luttinger exponent predicted by the SCHA and
    the renormalization group for $\alpha=5/2$ for variable $K$ and
    $2\pi J_{LR} A_0^2 a/u=0.1$. The SCHA predicts a lower value for
    the Tomonaga-Luttinger exponent than the RG. In particular, the
    SCHA prediction  does not reach the maximum value
    $1/(6-2\alpha)=1$. the SCHA also predicts instability of the TLL
    at a lower value of K.}
  \label{fig:scha-rg-alpha2.5}
\end{figure}

\begin{figure}[h]
  \centering
  \includegraphics[width=9cm]{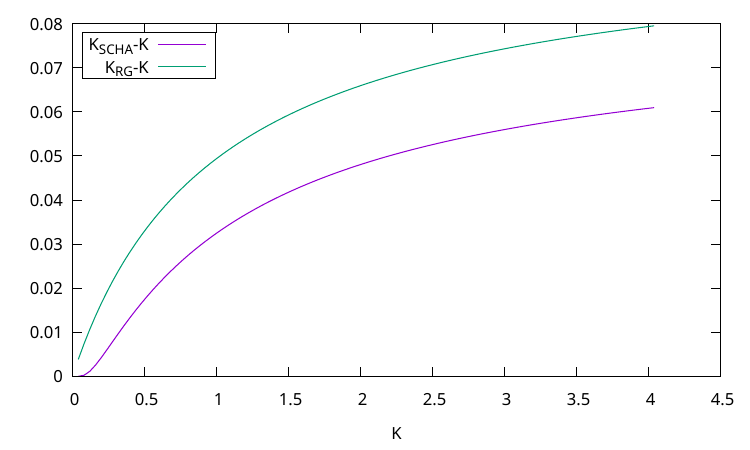} 
  \caption{Comparison of the Tomonaga-Luttinger exponent predicted by
     the SCHA (violet) and
    the renormalization group (green) for $\alpha=7/2$ for variable
    $K$ (bare Tomonaga-Luttinger exponent in the absence of long range hopping) and
    $2\pi J_{LR} A_0^2 a/u=0.1$. Given the smallness of the deviation from
    the bare exponent, the difference
    between bare exponent and exponent predicted by RG or SCHA has been
    plotted for clarity. The estimate from SCHA is always lower than
    the estimate from RG.}
  \label{fig:scha-rg-alpha3.5} 
\end{figure}

On Fig.~\ref{fig:scha-rg-alpha2.5}, we have plotted
$\tilde{K}$ and $K^*$ for $\alpha=5/2<3$. The SCHA predicts a lower
value of the exponent than the RG especially for $K \to K_c$. In
particular, the SCHA predicts a value of the Tomonaga-Luttinger
exponent below $1/(6-2\alpha)$ when $K\to K_c$. 
For $\alpha=7/2>3$, the predictions of SCHA and RG are represented on
Fig.~\ref{fig:scha-rg-alpha3.5}. The enhancement of the Tomonaga-Luttinger
exponent is smaller, and the SCHA prediction is lower than the RG
prediction. 

\subsection{Superfluid phase with broken symmetry}\label{sec:superfluid-scha}

\subsubsection{Dispersion relation and order parameter}
\label{sec:scha-sf-solution}

In the CSB phase, $\langle e^{i\theta}\rangle \ne 0$ and as a consequence,
\begin{equation}
  \label{eq:ginf-def}
  g(\infty)=\lim_{l \to +\infty} g(l)=|\langle e^{i\theta}\rangle|^2 >0. 
\end{equation}

For $ka \to 0$, Eq.~(\ref{eq:dispersion}) yields 
\begin{equation}\label{eq:omega-sf} 
  \omega(k)^2 \simeq u^2 k^2 +\frac{4 \pi J_{LR} A_0^2 u g(\infty) }{K a} \sum_{l=2}^{+\infty} \frac{1-\cos(kla)}{l^\alpha},  
\end{equation}
and with Eq.~(25.12.12) from Ref.~\onlinecite{olver2010nist}, we get
\begin{equation}
  \sum_{l=2}^{+\infty} \frac{1-\cos(kla)}{l^\alpha} =\frac{\pi}{2\Gamma(\alpha) \left|\cos \left(\frac{\pi\alpha} 2\right)\right|} (ka)^{\alpha-1} + O(ka)^2,   
\end{equation}
leading to $\omega(k)\sim (ka)^{(\alpha-1)/2}$ as $k\to 0$ for $1<\alpha<3$\cite{yusuf2004,laflorencie2005,maghrebi2017,vanderstraeten2018}. For $\alpha>2$, this result is compatible with a generalization\cite{liu2024,zhou2024,ma2024} of the Lieb-Schultz-Mattis theorem\cite{lieb_lsm_theorem} that predicts a gapless branch of excitations above a translationally invariant ground state. 
Since
\begin{equation}\label{eq:ginf-int}
  g(\infty)=\exp\left[-\int_0^{+\infty} \frac{d(uk)}{2 K \omega(k)} e^{-ka} \right].
\end{equation}
this makes the integral giving $g(\infty)$ nonzero, leading to selfconsistency.
So we can look for a CSB phase with $g(\infty)>0$ for $1<\alpha<3$,
 in the region where the Tomonaga-Luttinger liquid phase can
become unstable if interactions are not sufficiently repulsive.
The equation to consider is
\begin{eqnarray}
  \label{eq:selfconsistent-ginf}
  g(\infty)=\exp\left[-\frac {1} {2K} \int_0^{+\infty} \frac{d(uk)
  e^{-ka}}{\sqrt{(uk)^2 + \frac{2 \pi^2 J_{LR} A_0^2 u
  g(\infty)}{\Gamma(\alpha) |\cos(\pi\alpha/2)| K a} (ka)^{\alpha-1}}}
  \right]. \nonumber \\ 
\end{eqnarray}
By the change of variable $k=v^{2/(3-\alpha)}/a$, it is rewritten
\begin{eqnarray}
  g(\infty)&=&\exp\left[-\frac {1} {(3-\alpha)K} \int_0^{+\infty} \frac{dv e^{-v^{\frac{2}{3-\alpha}}}}{\sqrt{ \frac{2 \pi^2 J_{LR} A_0^2 g(\infty) a}{\Gamma(\alpha) |\cos(\pi\alpha/2)| u K } + v^2} }\right],\nonumber \\ 
\end{eqnarray}
For $g(\infty) \ll 1$, the integral is approximated as
\begin{equation}
  \ln \left[\frac{2 e^{-\frac{3-\alpha}2 \gamma_E}}{\sqrt{\frac{2\pi^2 J_{LR} A_0^2 g(\infty)}{\Gamma(\alpha) |\cos(\pi\alpha/2)| u K a}}}\right], 
\end{equation}
yielding
\begin{eqnarray}
  \label{eq:ginf-solution}
  g(\infty)=\left(\frac{\pi^2 J_{LR} A_0^2 a e^{(3-\alpha)\gamma_E}}{2\Gamma(\alpha) u K \left|\cos \left(\frac{\pi \alpha}{2}\right)\right|}\right)^{\frac{1}{2K(3-\alpha)-1}},
\end{eqnarray}
for $3-\alpha-(2K)^{-1}>0$ that is when $J_{LR}$ is relevant.
We can write $g(\infty)=(a/\xi)^{\frac{1}{2K}}$, with
\begin{eqnarray}\label{eq:corrlength-scha} 
  \xi=\left(\frac{\pi^2 J_{LR} A_0^2 a
  e^{(3-\alpha)\gamma_E}}{2\Gamma(\alpha) u K \left|\cos
  \left(\frac{\pi
  \alpha}{2}\right)\right|}\right)^{-\frac{1}{3-\alpha-\frac 1 {2K}}}. 
\end{eqnarray}
The lengthscale $\xi$ differs from $\xi_{\mathrm{RG}}$ defined in
Eq.~(\ref{eq:corrlength-rg}) only by a prefactor dependent on $K$ and
$\alpha$. The variation of $g(\infty)$ with $\xi$ corresponds with the
one expected from the scaling dimension of $\cos \theta$.
The SCHA is thus equivalent to the
vertical RG flow, in which the renormalization of $K$ is
neglected.\cite{suzumura_sg,citro04_spinpeierls}
In terms of the correlation length, the dispersion relation becomes
$\omega(k)=4K e^{(\alpha -3)\gamma_E} (u/\xi)^2 (k\xi)^{\alpha-1} + O(k^2)$ and
the terms of order $k^2$ are negligible for $k\xi \ll 1$. 
For the marginal case of $\alpha=3$, assuming $g(\infty)>0$, would
result in
\begin{eqnarray}
  \omega(k)^2=(uk)^2+\frac{2 J_{LR} A_0^2\pi u}{Ka} g(\infty) (ka)^2
  \ln(e^{1/2}/|k|a).\nonumber \\   
\end{eqnarray}
with the help of Eq.~(25.12.19) of \cite{olver2010nist}. However, the
integral in Eq.~(\ref{eq:ginf-int}) would then diverge, leading to
$g(\infty)=0$ in contradiction with the initial assumption. This rules
out long range ordering at $\alpha=3$. For $\alpha> 3$, we have
$\omega(k)^2 = O(k^2)$, so that $g(\infty)=0$ and only the TLL
solution can exist.

\subsubsection{Correlation functions in the CSB phase}
\label{sec:scha-sf-corr}

\paragraph{equal time correlations}

In the superfluid phase, the momentum distribution is given by
\begin{eqnarray}
  n(k)&=&A_0^2 \int dx e^{ik x} \langle e^{i\theta(x)} e^{-i\theta(x)}\rangle \nonumber \\
      &=& A_0^2 \left[2\pi g(\infty)\delta(k) + \int dx e^{ik x} (g(l)-g(\infty)) \right]\nonumber \\
  &=& A_0^2 \left[2\pi g(\infty) \delta(k) + \frac{g(\infty)}{4K\omega(k)}e^{-|k|a} +\ldots \right],   
\end{eqnarray}
yielding a momentum distribution with a power law divergence
$\sim (|k|\xi)^{-(\alpha-1)/2}$ for $ 0< |k|\xi \ll 1 $ and a Dirac
delta function peak at $k=0$ as in the noninteracting
  Bose gas. For $|k|\xi \gg 1$,
the TLL behavior $n(k) \sim (|k|a)^{\frac 1 {2K} -1}$ is recovered. 

The density-density correlation functions are obtained from
Eq.~(\ref{eq:bosonized-density}). First, let's consider the slowly varying
component.  According to Eq.~(\ref{eq:scha-fields}) 
\begin{eqnarray}
  \label{eq:sf-phase-density-q0}
  &&  \pi^{-2} \langle \partial_x \phi(x) \partial_x \phi(0)\rangle = \langle P(x) P(0)\rangle, \nonumber \\
  && =\frac{K}{\pi^2 a} \left[\frac{\pi^2 J_{LR} A_0^2
     e^{\frac{\gamma_E}{2K}} a}{2\Gamma(\alpha) \left|\cos
     \left(\frac{\pi \alpha}{2} \right)\right| u K}
     \right]^{\frac{(3-\alpha) K}{2K(3-\alpha)-1}} \nonumber \\
  && \times \int_0^{+\infty} dk (ka)^{(\alpha-1)/2} \cos(kx) e^{-ka}, \\
  &&=\frac{K^{3/2} e^{\gamma_E\frac{\alpha -3} 2 }}{2\pi \xi^2
     \Gamma\left(\frac{1-\alpha} 2 \right) \cos \frac \pi 4 (1-\alpha)
     } \left|\frac{\xi}{x}\right|^{(\alpha+1)/2} (|x|\gg \xi),  
\end{eqnarray}
with the correlation length $\xi$ given by
Eq.~(\ref{eq:corrlength-scha}). At long distances, density-density
correlations in the CSB phase decay as a power law, but more slowly
than in the TLL where they decay as inverse square of distance\cite{giamarchi_book_1d}.
Let's now turn our attention to the Fourier components of density-density
correlations
with wavevectors close to $2\pi n_0 m$. Using $\phi(k)=-i\frac{\pi P(k)} k$ and Eqs.~(\ref{eq:scha-fields}) we have
\begin{equation}
  \langle \phi(k) \phi(-k)\rangle =\frac{\pi K \omega(k)}{2u k^2},  
\end{equation}
and
\begin{equation}
  \frac 1 2 \langle (\phi(x)-\phi(0))^2\rangle = \frac K 4 \int_{-\infty}^{\infty} \frac{dk \omega(k) [1-\cos(kx)]}{2 u k^2}.  
\end{equation}
For long distance, that yields
\begin{equation}
  \langle (\phi(x)-\phi(0))^2\rangle =\frac{\pi
    K^{3/2}}{\Gamma\left(\frac{5-\alpha}{2}\right) \cos \frac \pi 4
    (1-\alpha)} \left(\frac {|x| e^{-\gamma_E}} \xi \right)^{\frac{3-\alpha}{2}},  
\end{equation}
and since 
\begin{eqnarray}
  \langle e^{2i m \phi(x)} e^{-2i m \phi(0)}\rangle = e^{-2 m^2  \langle (\phi(x)-\phi(0))^2\rangle},   
\end{eqnarray}
the density-density correlation functions at wavevector $2\pi m n_0$
decay as stretched exponentials\cite{maghrebi2017}. The final result
for the density-density correlation is 
\begin{widetext}
\begin{eqnarray}
  \langle n(x) n(0)\rangle = \frac{K^{3/2}
  e^{\gamma_E\frac{\alpha -3} 2 }} {2\pi \xi^2
     \Gamma\left(\frac{1-\alpha} 2 \right) \cos \frac \pi 4 (1-\alpha)
     } \left|\frac{\xi}{x}\right|^{(\alpha+1)/2} +
  \sum_{m=1}^{+\infty} \frac{B_m^2}{2}  e^{- \frac{2 m^2 \pi
    K^{3/2}}{\Gamma\left(\frac{5-\alpha}{2}\right) \cos \frac \pi 4
    (1-\alpha)} \left(\frac {|x| e^{-\gamma_E}} \xi
  \right)^{\frac{3-\alpha}{2}}} \cos (2\pi m n_0 x/a) 
\end{eqnarray}
\end{widetext}
where $n(ja)=(b^\dagger_j b_j -n_0)/a$. 
The static structure factor is obtained by Fourier transformation of
the above expression. In the vicinity of $q=0$, it behaves as $\sim
|q|^{(\alpha -1)/2}$,
while in the vicinity of $q=2 \pi m n_0$, it can be expressed in terms of Fox $H$-functions\cite{hilfer2002,mathai2010}.

\paragraph{Time dependent correlations}

The time dependent correlation functions are
\begin{widetext} 
\begin{eqnarray}
  \langle T_\tau \theta(x,\tau) \theta(0,0) \rangle &=&
  \frac{u}{2\bar{u} K} \int_0^{+\infty} \frac{dv}{v^{(\alpha-1)/2}}
  \cos \left(\frac{v x}{\xi}\right) e^{-\frac{\bar{u} \tau}{\xi}
                                                        v^{(\alpha-1)/2}}, \nonumber \\
  &=&  \frac{u}{2\bar{u} K} \mathrm{Re}\left\{ e^{i\pi \frac {3-\alpha} 4}\left(\frac \xi
      {|x|}\right)^{\frac {3-\alpha} 2}
      {}_1\Psi_0\left[-\frac{\bar{u} |\tau|}{\xi}\left(\frac \xi
      {|x|}\right)^{\frac {\alpha-1} 2} e^{i\pi \frac {\alpha-1} 4}\left|\left(\frac{3-\alpha} 2, \frac{\alpha -1}
      2\right) \right. \right]\right\},
\end{eqnarray}
\end{widetext}
where ${}_1\Psi_0$ is a Fox-Wright hypergeometric function\cite{mathai2010}. The
correlator satisfies scaling with a dynamical exponent
$(\alpha-1)/2$. For $x=0$, it decays as
$|\tau|^{-\frac{3-\alpha}{1-\alpha}}$, so that
\begin{equation}
    \lim_{\tau \to \infty} \langle e^{i\theta(0,\tau)}
    e^{-i\theta(0,0)} \rangle =
    |\langle e^{i\theta}\rangle|^2.
\end{equation}
For density-density correlations, we have
\begin{widetext} 
\begin{eqnarray} 
\frac 1 {\pi^2}  \langle T_\tau \partial_x \phi(x,\tau) \partial_x \phi(0,0) \rangle &=&
  \frac{\bar{u} K }{2\pi^2 u \xi^2} \int_0^{+\infty} dv v^{(\alpha-1)/2} 
  \cos \left(\frac{v x}{\xi}\right) e^{-\frac{\bar{u} \tau}{\xi}
                                                        v^{(\alpha-1)/2}}, \nonumber \\
  &=&   \frac{\bar{u} K }{2\pi^2 u \xi^2}  \mathrm{Re}\left\{ e^{i\pi \frac {\alpha+1} 4}\left(\frac \xi
      {|x|}\right)^{\frac {\alpha+1} 2}
      {}_1\Psi_0\left[-\frac{\bar{u} |\tau|}{\xi}\left(\frac \xi
      {|x|}\right)^{\frac {\alpha-1} 2} e^{i\pi \frac {\alpha-1} 4}\left|\left(\frac{\alpha+1} 2, \frac{\alpha -1}
      2\right) \right. \right]\right\},
\end{eqnarray}
\end{widetext}
with a decay as $|\tau|^{-\frac{\alpha+1}{\alpha-1}}$ for $x=0$, and
\begin{eqnarray}
&&\frac 1 2 \langle T_\tau[\phi(x,\tau)-\phi(0,0)]^2 \rangle =
    \frac{\bar{u} K}{2u} \int_0^{+\infty} dv v^{(\alpha-5)/2}\times \nonumber \\ 
&&   \left[1-\cos \left(v \frac x \xi \right) e^{-\frac{\bar{u}
    |\tau|}{\xi} v^{(\alpha-1)/2}} \right].     
\end{eqnarray}
  The latter integral is divergent when $\alpha<2$ and $\tau\ne
  0$. In the case of the quantum sine-Gordon model, it is known that
  the SCHA gives a similar divergence when calculating the correlation
  functions of the dual field\cite{watanabe_2ch} as it overestimates
  the energy cost of propagating solitons in
  time\cite{lukyanov_soliton_ff}. So the divergence might be an
  artefact of the SCHA. For $\alpha>2$, no divergence exists, and 
\begin{widetext}
  \begin{eqnarray}
    \frac 1 2 \langle T_\tau[\phi(x,\tau)-\phi(0,0)]^2 \rangle =
    - \frac{\bar{u} K}{2u}\left(\frac{|x|}{\xi}\right)^{\frac{3-\alpha}2} \mathrm{Re}\left\{{}_1\Psi_0\left[-\frac{\bar{u} |\tau|}{\xi}\left(\frac \xi
      {|x|}\right)^{\frac {\alpha-1} 2} e^{i\pi \frac {\alpha-1} 4}\left|\left(\frac{\alpha-3} 2, \frac{\alpha -1}
      2\right) \right. \right]\right\},  
  \end{eqnarray}
\end{widetext}
with growth as $|\tau|^{\frac{3-\alpha}{\alpha-1}}$ as $\tau \to
+\infty$. When $\alpha>2$, density-density correlations with
wavevector near $2\pi m n_0$ show a stretched exponential decay with
Matsubara time.

\subsection{Frustrating long range interaction}
\label{sec:frustrating}

If we consider the XY antiferromagnetic spin chain in the presence of the long-range antiferromagnetic exchange interaction
\begin{equation}
  H_{LR}^{\mathrm{frus.}} = - J_{LR}^f \sum_j \sum_{l=2}^{+\infty}
  \frac{1}{l^\alpha} (S^+_j S^{-}_{j+l}+S^-_j S^{+}_{j+l}),  
\end{equation}
no matter the sign of $J_{LR}^f$, that interaction is
incompatible with the antiferromagnetic ordering favored by the
nearest-neighbor interaction, leading to frustration in the model. 
Applying bosonization  yields a perturbation
\begin{equation}
  \label{eq:frustrated-boso}
H_{LR}^{\mathrm{frus.}} =-\frac{2J_{LR}^f A_0^2}{a} \int dx \sum_l
\frac{(-1)^l}{l^\alpha}  \cos [\theta(x+la)-\theta(x)]. 
\end{equation}
It is more difficult to turn Eq.~(\ref{eq:frustrated-boso}) into an
integral to apply a RG treatment, but the SCHA treatment remains
simple. 
Using the SCHA, our Hamiltonian becomes identical to
Eq.~(\ref{eq:scha-fourier}), albeit with $g(l) \to (-1)^l g(l)$. If we
look for a Tomonaga-Luttinger liquid solution, we have
\begin{widetext}
\begin{eqnarray}
  \label{eq:scha-frustrated}
   \mathcal{H}_{SCHA}&=&\sum_k \left[\frac {\pi u}{2K} P(k) P(-k) +
                         \left(\frac{uK k^2}{2\pi} + \frac{2 J^f_{LR}
                         A_0^2}{a} [\mathrm{Li}_{\alpha+\frac 1
                         {2\tilde{K}}} (-1) -\mathrm{Re} (\mathrm{Li}_{\alpha+\frac 1
                         {2\tilde{K}}} (-e^{ika})) + 1-\cos (ka)]
                         \right)  \theta(k) \theta(-k) \right],
                         \nonumber \\  
\end{eqnarray}
\end{widetext}
where $\mathrm{Li}_{\alpha+\frac 1 {2\tilde{K}}}$ is the
  polylogarithm\cite{olver2010nist}. After expanding to second order
  in $k$, we obtain $u/K=\tilde{u}/\tilde{K}$ and
  \begin{eqnarray}\label{eq:scha-frustrated-TLL} 
    \tilde{K}^2 = K^2 +\frac{2\pi J_{LR}^f K A_0^2 a}{u}
    \left[1-\int_0^{+\infty} \frac{dv v^{\alpha+\frac 1{2\tilde{K}}
    -1}}{\Gamma\left(\alpha+\frac 1{2\tilde{K}}\right)}
    \frac{e^{2v}-e^v}{(e^v+1)^3} \right]. \nonumber \\  
  \end{eqnarray}
  In contrast to the unfrustrated case, Eq.~(\ref{eq:scha-TLL}),
  letting $\tilde{K}$ to infinity in the right hand side of
  Eq.~(\ref{eq:scha-frustrated-TLL}), gives a finite expression for
  any $\alpha>0$. Using Eq.~(25.5.21) in
  Ref.~\onlinecite{olver2010nist}, we can express the right hand side
  of Eq.~(\ref{eq:scha-frustrated-TLL}) using an analytically
  continued Riemann zeta function.
  Now let's try to find a continuum limit for
  Eq.~(\ref{eq:frustrated-boso}) by introducing the sums 
  \begin{equation}
    \label{eq:staggered-sum}
    S_j=\sum_{l=-\infty}^j (-1)^j e^{i\theta(ja)},  
  \end{equation}
  and rewriting
  \begin{eqnarray}
    H_{LR}^{\mathrm{frus.}} &=&- J_{LR}^f A_0^2 \sum_j \sum_k
    \frac{1}{|j-k|^\alpha}
                                [(S_j-S_{j-1})(S_k-S_{k-1})^\dagger\nonumber
    \\
    &&+ (S_j-S_{j-1})^\dagger(S_k-S_{k-1})] \nonumber \\
    &=& - J_{LR}^f A_0^2 \sum_j \sum_k [S_j S_k^\dagger+S_j^\dagger
        S_k]\times \nonumber \\
    && \left[\frac{2}{|j-k|^\alpha} - \frac{1}{|j+1-k|^\alpha} -
        \frac{1}{|j-1-k|^\alpha}\right].    
  \end{eqnarray}
  If we now write
  \begin{eqnarray}
    S_{2j}&=&e^{i \theta(2ja)} -e^{i \theta(2ja-a)}+ e^{i
    \theta((2j-2)a)} -e^{i \theta((2j-3)a)}+\ldots \nonumber \\
    &\simeq& \int_{-\infty}^{2 ja} dx \frac{d}{dx} (e^{i\theta})
             \nonumber \\
    &\simeq& e^{i\theta(2ja)},               
  \end{eqnarray}
  and $S_{2j+1} =S_{2j}-e^{i \theta(2ja+a)}\simeq 0$, and we note that
  $2 j^{-\alpha} -(j+1)^{-\alpha} -(j-1)^{-\alpha}=\alpha(\alpha+1)
  j^{-\alpha-2}+o(j^{-\alpha-2})$, we find an unfrustrated interaction
  \begin{equation} 
    H_{LR}^{\mathrm{frus.}} \sim - J_{LR}^f A_0^2 \int dx dy
    \frac{\cos (\theta(x) -\theta(y))}{|x-y|^{\alpha+2}}. 
  \end{equation}
  The renormalization group treatment can proceed along the lines of
  Sec.~\ref{sec:rg} by replacing in Eq.~(\ref{eq:rg-reduced}) $\alpha$
  with $\alpha+2$. The resulting interaction $g_{LR}$ is always
  irrelevant, leading to a stable TLL phase in the frustrated case. 
  
  A closely related problem is the effect of long range interactions
or easy axis exchanges at incommensurate filling. 
If we consider the latter, 
\begin{equation}
  \label{eq:lr-interaction}
H_{LR}^z=  J^z_{LR} \sum_{j} \sum_{l=2}^{+\infty} \frac {S^z_{j+l} S^z_j}{l^\alpha},  
\end{equation}
bosonization will yield
\begin{eqnarray}
  \label{eq:lr-interact-boso}
H_{LR}^z&=& \int dx  \sum_{l=2}^{+\infty} \frac{J_{LR}^z}{a \pi^2 l^\alpha} \partial_x\phi(x)
            \partial_x \phi(x+la)  \\
  && +\int dx \sum_{l=2}^{+\infty} \frac{J_{LR}^z B_1^2}{2 a l^\alpha} \cos (2\phi(x+la) -2
     \phi(x) -2k_F l a), \nonumber 
\end{eqnarray}
where we have dropped the contribution $\propto \cos (2\phi(x+la) -2
     \phi(x) +2k_F l a + 4k_F x)$ since $k_F a \ne \pi /2$.    
In Fourier representation, the forward scattering interaction in the
first line of Eq.~(\ref{eq:lr-interact-boso}), takes the form
\begin{eqnarray}
  \sum_k \frac{J^z_{LR}}{\pi^2} \sum_{l=2}^{+\infty}  \frac{\cos(k la)}{l^\alpha} k^2
  \phi(k) \phi(-k),  
\end{eqnarray}
and provided $\alpha>1$, can be approximated\cite{inoue_conformal_2006} as
\begin{equation}
  \frac{J_{LR}^z (\zeta(\alpha) -1) }{\pi^2 a } \int dx (\partial_x
  \phi)^2.   
\end{equation}
In the case $\alpha=1$, a logarithmic factor is
present.\cite{schulz_wigner_1d}  
With a magnetized spin chain, $k_F \ne \pi/(2a)$ and we find a
slightly more general form of the SCHA 
\begin{eqnarray}
  H_{SCHA}&=&\sum_k \left[\frac{\pi uK} 2 \Pi(k) \Pi(-k) + \frac {uK
              k^2 }{2\pi} \phi(k) \phi(-k) \right] \nonumber \\
          &&+ \sum_k \frac{J_{LR}^z (\zeta(\alpha) -1) k^2 }{\pi^2 a }
             \phi(k) \phi(-k) \nonumber \\ 
          && -\frac{2J_{LR}^z B_1^2}{a} \sum_k
             \left[\mathrm{Re}(\mathrm{Li}_{\alpha+2\tilde{K}}(e^{2ik_F
             a})) \right. \nonumber \\
          && \left. - \frac 1 2
             \mathrm{Re}(\mathrm{Li}_{\alpha+2\tilde{K}}(e^{i (2k_F -k)
             a})) \right. \nonumber \\
          && \left. - \frac 1 2
             \mathrm{Re}(\mathrm{Li}_{\alpha+2\tilde{K}}(e^{i (2k_F +k)
             a})) \right. \nonumber \\
          &&  \left. -\cos (2k_F a) (1-\cos (ka)) \right]
             \phi(k)\phi(-k).                                                  
\end{eqnarray}
The expression with polylogarithms in the last line has the Taylor
expansion
\begin{widetext} 
\begin{eqnarray}
 && \mathrm{Re}\left[ \mathrm{Li}_{\alpha+2\tilde{K}}(e^{2ik_F
             a})- \frac 1 2 \mathrm{Li}_{\alpha+2\tilde{K}}(e^{i (2k_F -k)
             a}) + \frac 1 2 \mathrm{Li}_{\alpha+2\tilde{K}}(e^{i (2k_F -k)
    a}) \right] \nonumber \\
  && = -\frac{(ka)^2}{4\Gamma(\alpha+2\tilde{K})}
  \int_0^{+\infty} dv v^{\alpha+2\tilde{K}-1} \sinh v \frac{\cos (2k_F
  a) (\cosh v - \cos(2k_F a))+ 2\sin^2(2k_F a)}{(\cosh v -\cos (2k_F
  a))^3} +o(k^2), 
\end{eqnarray}
\end{widetext}
which is finite for $\alpha>1$ and $k_F \ne 0$. So the
Tomonaga-Luttinger liquid is preserved by long range easy axis
exchange in a magnetized spin chain. 

\subsection{Effect of temperature}
\label{sec:temperature}
Till now, we have been considering only the ground state. In the
present section, we consider the stability of the CSB phase to thermal
fluctuations, and the effect of thermal fluctuations in the
Tomonaga-Luttinger exponent. 
\subsubsection{Stability of the CSB phase} 
When $T>0$, Eq.~(\ref{eq:scha-g-l}) is replaced by
\begin{equation}
  \label{eq:scha-g-l-temp}
  g(l)=\exp\left[-\int_0^{\infty} \frac{d(uk) e^{-ka}}{2K\omega(k)} \left(1+\frac{2}{e^{\frac{\omega(k)}T} -1}\right)(1-\cos(kla)) \right]. 
\end{equation}
In order to have a CSB solution with $g(\infty)>0$, we need to solve 
\begin{equation}\label{eq:ginf-scha-temp} 
  g(\infty)=\exp\left[-\int_0^{\infty} \frac{d(uk) e^{-ka}}{2K\omega(k)} \left(1+\frac{2}{e^{\frac{\omega(k)}T} -1}\right) \right].
\end{equation}
with $\omega(k)$ given by Eq.~(\ref{eq:omega-sf}). To find
$g(\infty)>0$, the integral 
\begin{equation}
  T\int_0^{1/a} \frac{dk}{\omega(k)^2} \sim T\int_0^{1/a}
  \frac{dk}{k^{\alpha -1}},  
\end{equation}
must be finite. Thus, for $\alpha>2$, CSB disappears as soon as $T>0$, in agreement with Ref.~\cite{bruno2001}. 
When $\alpha <2$, we can introduce a temperature dependent correlation length by
\begin{equation}
  \frac a {\bar{\xi}(T)} = \left(\frac{2 \pi^2 J_{LR} A_0^2 a g(\infty) }{\Gamma(\alpha) u K \left|\cos \left(\frac{\pi \alpha}{2}\right)\right|}\right)^{\frac{1}{3-\alpha}},
\end{equation}
such that $\omega(k)^2=u^2
[(k\bar{\xi})^2+(k\bar{\xi})^{\alpha-1}]/\bar{\xi}^2$. The amplitude
of the dispersion becomes temperature dependent. We rewrite
Eq.~(\ref{eq:ginf-scha-temp}) in the form
\begin{eqnarray}
  g(\infty)=\exp\left[-\int_0^{+\infty}
  \frac{d\lambda e^{-\lambda a/\xi} }{2K \sqrt{\lambda^{\alpha-1}+\lambda^2}}  \cotanh
  \left(\frac{\beta u}{2\xi}\sqrt{\lambda^{\alpha-1}+\lambda^2}
  \right)\right],  \nonumber \\ 
\end{eqnarray}
At sufficiently low temperature, we split the positive axis into three
regions, $0<\lambda < (2T \xi/u)^{2/(\alpha-1)}$,
$ (2T \xi/u)^{2/(\alpha-1)} < \lambda <1$ and $\lambda>1$, and
approximate the integrand on each of them. We find
\begin{eqnarray}
&&  -\ln g(+\infty) \simeq \frac{\xi}{\beta u K}
  \int_0^{\left(\frac{2T\xi}{u}\right)^{\frac{2}{\alpha-1}}}
   \frac{d\lambda}{\lambda^{\alpha-1}} \nonumber \\
  && + \frac 1 {2K}
  \int_{\left(\frac{2T\xi}{u}\right)^{\frac{2}{\alpha-1}}}^1
     \frac{d\lambda}{\lambda^{\frac{\alpha-1} 2}}  + \int_1^{+\infty}
  \frac{d\lambda}{\lambda} e^{-\lambda a /\xi} 
\end{eqnarray}
The selfconsistent equation (\ref{eq:ginf-scha-temp}) becomes 
\begin{eqnarray}\label{eq:scha-temp-approx} 
&&  \left(\frac{a}{\bar{\xi}(T)}\right)^{3-\alpha-\frac 1 {2K}}  =
   \frac{2 \pi^2 J_{LR} A_0^2 a }{\Gamma(\alpha) u K \left|\cos
   \left(\frac{\pi \alpha}{2}\right)\right|} e^{-\frac{1}{(3-\alpha)
   K}} \nonumber \\
  && \times \exp\left[-\frac {\alpha -1}{2K(2-\alpha) (3-\alpha)} \left(\frac{2T \bar{\xi}(T) }u\right)^{\frac{3-\alpha}{\alpha-1}}\right],  
\end{eqnarray}
which, for $T=0$,  reduces  to
\begin{equation}
  \left(\frac{a}{\bar{\xi}(T=0)}\right)^{3-\alpha-\frac 1 {2K}}  =  \frac{2 \pi^2 J_{LR} A_0^2 a e^{\frac{\gamma_E}{2K}}}{\Gamma(\alpha) u K \left|\cos \left(\frac{\pi \alpha}{2}\right)\right|}  e^{-\frac{1}{(3-\alpha) K}}.  
\end{equation}
This expression differs from (\ref{eq:corrlength-scha}) by prefactors
dependent on $K$ and $\alpha$ because we have used a cruder
approximation than in Sec.~\ref{sec:superfluid-scha} to approximate
the finite temperature integral. 
For temperatures $T<T_c$ with
\begin{eqnarray}\label{eq:tcrit-scha} 
 && T_c =\frac{u}{2a} \left[\frac{2 \pi^2 J_{LR} A_0^2 a
    e^{\frac{\gamma_E}{2K}}}{\Gamma(\alpha) u K \left|\cos
        \left(\frac{\pi \alpha}{2}\right)\right|}
    e^{-\frac{1}{(3-\alpha)
  K}}\right]^{\frac{1}{3-\alpha-\frac{1}{2K}}}  \nonumber \\
  && \times \left[\frac{(2-\alpha)(2K(3-\alpha)-1)}{e}\right]^{\frac{\alpha-1}{3-\alpha}}, 
\end{eqnarray}
the correlation length behaves as 
\begin{eqnarray}
 && \bar{\xi}(T)=\frac{u}{2T}
 \left[(2-\alpha)(2K(3-\alpha)-1)\right]^{\frac{\alpha-1}{3-\alpha}}
 \nonumber \\
 && \times \left|W_P\left(-\frac 1 e \left(\frac{T}{T_c}\right)^{\frac{3-\alpha}{\alpha-1}} \right)\right|^{\frac{\alpha-1}{3-\alpha}},   
\end{eqnarray}
where $W_p$ is the principal branch of the Lambert W
function\cite{olver2010nist}. According to Eq.~(\ref{eq:tcrit-scha}),
the critical temperature vanishes when $\alpha \to 2$ as expected. The
temperature dependence for the square modulus of the CSB order parameter is given by 
\begin{eqnarray}\label{eq:ginf-t-scha} 
  g(\infty)&=&\left(\frac {a}{\bar{\xi}(T=0)}\right)^{\frac 1 {2K}}
  e^{-\frac{1}{(3-\alpha)K}} \nonumber \\
  && \times\exp \left[(\alpha -1) W_P\left(-\frac 1
        e \left(\frac{T}{T_c}\right)^{\frac{3-\alpha}{\alpha-1}}
      \right)\right]. 
\end{eqnarray}

\begin{figure}[h]
  \centering
  \includegraphics[width=9cm]{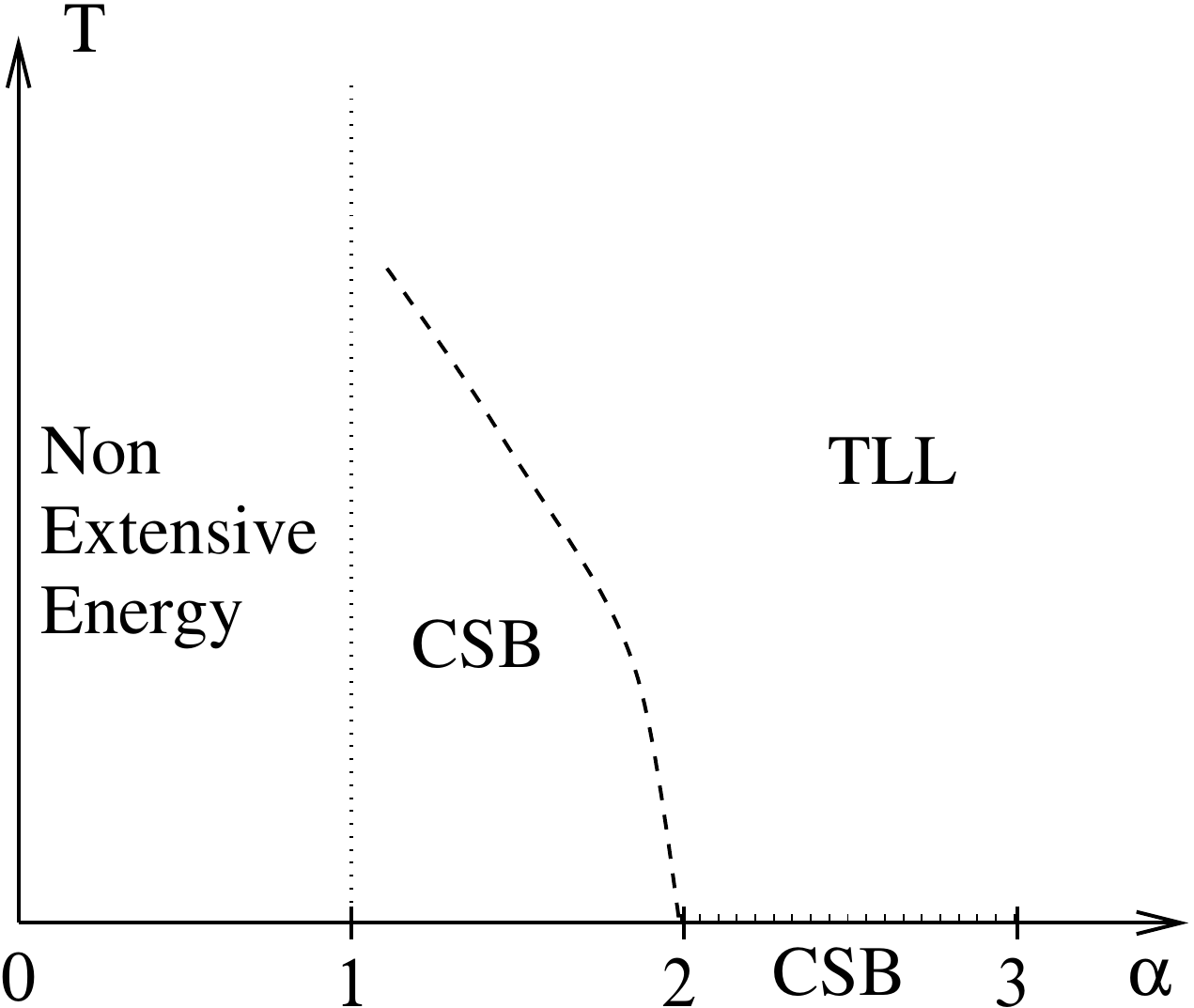} 
  \label{fig:phasediag}
  \caption{Phase diagram in the $(\alpha,T)$ plane. For $0<\alpha<1$, the energy is non-extensive. For $1<\alpha<2$ a phase with continuous symmetry breaking (CSB) exists below the critical temperature represented by the dashed line. At higher temperature, the Tomonaga-Luttinger liquid (TLL) is recovered. For $2\le \alpha <3$, the CSB is stable only in the ground state. For $\alpha\ge 3$, the TLL is stable at all temperature.  }
\end{figure}

For $T>T_c$ the solution is $\bar{\xi}(T)=+\infty$ and
$g(\infty)=0$. Right at $T=T_c$ Eq.~(\ref{eq:ginf-t-scha}) gives
$g(\infty)=[a/\bar{\xi}(T=0)]^{1/(2K)} e^{1-\alpha -1/[K(3-\alpha)]}$.
The discontinuity in $g(\infty)$ is a well known artifact of the
SCHA\cite{donohue_thesis}. At finite
temperature, the classical one-dimensional XY model with long range interaction
 has a continuous phase transition\cite{fisher1972,takamoto_2010} for
$\alpha<2$, with mean-field behavior with $\alpha<3/2$, so we expect
a continuous transition also in the quantum model.  
The resulting phase diagram is sketched on Fig.~\ref{fig:phasediag}
\subsubsection{Temperature dependence of the Tomonaga-Luttinger exponent} 
In the Tomonaga-Luttinger liquid phase, the velocity and the Luttinger exponent are still given by Eqs.~(\ref{eq:uoverK})--~(\ref{eq:utimesK}), but with
\begin{equation}
  \label{eq:g-l-tll-temp}
  g(l)=\left(\frac{\pi a}{\beta \tilde{u} \sinh \frac{\pi l a}{\beta \tilde{u}}} \right)^{\frac 1 {2\tilde{K}}}, 
\end{equation}
where $\beta=1/T$ so that
\begin{equation}\label{eq:K-temp} 
  \tilde{u} \tilde{K}=  u K  + 2\pi J_{LR} A_0^2 a
  \sum_{l=2}^{+\infty} l^{2-\alpha} \left(\frac{\pi T a}{ \tilde{u}
      \sinh \frac{\pi T l a}{ \tilde{u}}} \right)^{\frac 1
    {2\tilde{K}}},   
\end{equation}
and $\tilde{u}/\tilde{K}=u/K$. 
At high temperature, $T \gg u/a$, $\tilde{K}=K$. As temperature is
lowered, $\tilde{K}$ increases. If $3-\alpha-(2\tilde{K})^{-1}<0$,
the ground state value is recovered when $T\to 0$. In the opposite
case,  the Tomonaga-Luttinger
exponent $\tilde{K}$ diverges when $T\to 0$. The leading correction in
Eq.~(\ref{eq:K-temp}) yields (see App.~\ref{app:summation})
\begin{eqnarray}
 \tilde{u} \tilde{K}=  u K+ 2 \pi  J_{LR} A_0^2 a \Gamma(3-\alpha) 
  \left(\frac{2 \tilde{u} \tilde{K}}{\pi T a}\right)^{3-\alpha} (1+o(1)),  
\end{eqnarray}
so that, neglecting $uK$, 
\begin{equation}
 \tilde{K} = \left[\frac{2\pi K \Gamma(3-\alpha) J_{LR} A_0^2
     a}{u}\right]^{\frac{1}{2\alpha-4}} \left(\frac{2 u}{\pi T a
     K}\right)^{\frac{3-\alpha}{2\alpha-4}},   
\end{equation}
when $2<\alpha<3$. The TLL exponent diverges as a power law, only when
$T\to 0$.
For $\alpha=2$, we approximate the sum for $\tilde{K}\gg 1$ by
replacing the hyperbolic sine with an exponential. We find 
\begin{equation}
  \tilde{u}\tilde{K}=u K + 2\pi J_{LR} A_0^2 a \frac{e^{-\frac{\pi T a}{\tilde{u}\tilde{K}}}}{1-e^{-\frac{\pi T a}{2\tilde{u}\tilde{K}}}}
\end{equation}
We can obtain the temperature as a function of $\tilde{u} \tilde{K}$
from that expression. We find that the expression has a minimum as a
function of $\tilde{u} \tilde{K}$ indicating that the
Tomonaga-Luttinger liquid becomes unstable below a certain temperature
$T^*$. Since we have seen previously that for $\alpha=2$,
there was no long range order for $T>0$, we can conjecture that, as in
the classical case\cite{takamoto_2010}, a quasi-long range order is
obtained for $0<T<T^*$. Unfortunately, for $\alpha=2$, the SCHA only
describes the TLL phase and cannot predict a phase with quasi-long
range order. More refined approximations will be needed to tackle this
question.   
When $\alpha<2$, we find that
\begin{equation}
  1=\frac{K^2}{\tilde{K}^2} + \frac{2 \pi K J_{LR} A_0^2 a}u
  \Gamma(3-\alpha) \left(\frac{u}{\pi a T K}\right)^{3-\alpha}
  \tilde{K}^{4-2\alpha}. 
\end{equation}
The expression in the right hand side has a minimum equal to
\begin{eqnarray}
 &&  \frac{2u K}{\pi a T(2-\alpha) } \left(\frac{2\pi J_{LR} A_0^2 a}{u
      K}\right)^{\frac 1 {3-\alpha}}
  \left[\frac{1}{\Gamma(3-\alpha)} \right. \\
  && \left. + [(2-\alpha)
      \Gamma(3-\alpha)]^{3-\alpha}\right],  
  \end{eqnarray}
  and for sufficiently low temperature, no solution is possible,
  indicating the instability of the TL solution as we have found for $\alpha=2$.
  The transition temperature thus obtained behaves as $\sim (J_{LR}
  a/u)^{1/(3-\alpha)}$, in disagreement with
  Eq.~(\ref{eq:tcrit-scha}). Moreover, the TLL exponent remains
  finite at the transition instead of diverging: the SCHA does not describe correctly the
  critical behavior near the transition point coming from high
  temperature. 
\subsubsection{Density-density correlations}

For $T>0$, we have 
\begin{equation}
  \langle \phi(k) \phi(-k)\rangle =\frac{\pi K \omega(k)}{2u k^2}\left[1+\frac{2}{e^{\omega(k)/T} -1} \right],  
\end{equation}
and
\begin{eqnarray}
&&  \frac 1 2 \langle (\phi(x)-\phi(0))^2\rangle = \frac K 4
  \int_{-\infty}^{\infty} \frac{dk \omega(k) [1-\cos(kx)]}{2 u k^2}
  \nonumber \\
  && \times \left[1+\frac{2}{e^{\omega(k)/T} -1} \right].  
\end{eqnarray}
In the superfluid phase with $\alpha<2$ and at long distance, this yields
\begin{equation}
  \langle (\phi(x)-\phi(0))^2\rangle =\frac{\pi K}{\Gamma\left(\frac{5-\alpha}{2}\right) \cos \frac \pi 4 (1-\alpha)} \left(\frac x \xi \right)^{\frac{3-\alpha}{2}} +\frac{\pi T|x|}{2u},  
\end{equation}
so that the zero-temperature stretched exponential is now multiplied by $e^{-m^2 \pi T|x|/u}$. In the Tomonaga-Luttinger liquid phase, the usual expression\cite{giamarchi_book_1d} is recovered.

\section{Conclusion}
\label{sec:disc-concl}

In conclusion, we have considered the effect of long range hopping or long range
interaction decaying as power law with distance in the Bose Hubbard
model using bosonization combined with the renormalization group method and the self-consistent Hamonic
approximation. Since a spin-1/2 chain with easy-plane or easy-axis echange
interaction can be mapped to hard core bosons using the Matsubara-Matsuda\cite{matsubara1956}
transformation, the conclusions are also relevant for atiferromagnetic
spin-1/2 chains with long range easy-plane or easy-axis exchange.
In the case of unfrustrated hopping  decaying more slowly than the cube of
the distance, a continuous symmetry breaking ground state is
stabilized when on site repulsion is weak enough. In such
state, the excitations have a power law dispersion,\cite{yusuf2004,laflorencie2005} and the
density-density correlations decay as stretched exponentials\cite{maghrebi2017} with
distance. 
With interchain hopping decays more rapidly than the cube of the
distance, or strong enough short
distance  repulsion, the Tomonaga-Luttinger liquid
ground state is stable.  When long range hopping is frustrated, the
Tomonaga-Luttinger liquid is always stable\cite{maghrebi2017}. At positive temperature,
the continuous symmetry breaking (CSB) phase is stable only when the
unfrustrated hopping decays more slowly than the square of the
distance. For faster decay, a Tomonaga-Luttinger liquid (TLL) is
stabilized, with a Tomonaga-Luttinger exponent diverging at zero
temperature. 
The case of long range interactions at incommensurate filling is
analogous to the case of frustrated hopping, with only a
Tomonaga-Luttinger liquid phase in the ground state. In the case of commensurate filling,
a long range ordering is expected to compete with the
Tomonaga-Luttinger liquid\cite{giamarchi_book_1d}.
Various open questions remain. The SCHA predicts a discontinuous
transition between the CSB and the TLL at positive
temperature, while the transition is actually of second
order\cite{fisher1972,takamoto_2010}. It would be interesting to determine its critical
exponents, in particular the dynamical exponent $z$, and the behavior
of the Tomonaga-Luttinger exponent at the transition. For unfrustrated
hopping decaying as inverse square of the distance, the SCHA predicts
instability of the Tomonaga-Luttinger liquid below a critical
temperature. In the classical system, a BKT-like transition to a phase
with quasi-long range order is expected\cite{takamoto_2010}. The SCHA
fails to capture a quasi-long range order at low temperature. Further
studies are necessary to determine whether a quasi-long range ordered
phase can be present with inverse square hopping. 

\begin{acknowledgments}
I thank T. Roscilde, F. Mezzacapo, S. Bocini, F. Caleca for
discussions.   
\end{acknowledgments}
\appendix
\section{Derivation of the renormalization group equations with
  operator product expansion}\label{app:ope-rg}
We start from the action~(\ref{eq:ls-boso-contin}), and we consider
the effect of increasing the cutoff from $a$ to $ae^{d\ell}$. Under
the increase of the cutoff, the double integral on $x$ and $y$
contributes
\begin{widetext}
\begin{eqnarray}
  -\frac{J_{LR} A_0^2}{a^{2-\alpha}} \left\{\int d\tau \int_{a\le |x-y|\le
  a e^{d\ell}} \frac{\cos
  [\theta(x,\tau)-\theta(y,\tau)]}{|x-y|^\alpha} +\int d\tau \int_{ |x-y|\ge
  2a e^{d\ell}} \frac{\cos
  [\theta(x,\tau)-\theta(y,\tau)]}{|x-y|^\alpha} \right\}.    
\end{eqnarray}
\end{widetext}
In each integral over the intervals $a\le |x-y|\le a e^{d\ell}$, we
 apply the OPE
\begin{eqnarray}
  \cos [\theta (x,\tau) -\theta (y,\tau)] = 1 - \frac{(x-y)^2}{2}
  (\partial_x \theta)^2 + \ldots,   
\end{eqnarray}
and rewrite their contribution to the action in the form
\begin{eqnarray}
  -\frac{J_{LR} A_0^2}{a^{2-\alpha}} \int d\tau dx \int_{x-a
  e^{d\ell}}^{x-a} dy \frac { 1 - (x-y)^2
  (\partial_x \theta)^2/2 + \ldots} {|x-y|^\alpha}  \nonumber \\
  -\frac{J_{LR} A_0^2}{a^{2-\alpha}} \int d\tau dx \int_{x+a}^{x+a
  e^{d\ell}} dy \frac {1 - (x-y)^2
  (\partial_x \theta)^2/2 + \ldots}{|x-y|^\alpha} , \nonumber \\
\end{eqnarray}
resulting in a net contribution
\begin{eqnarray}
 \int dx d\tau \left[ -2 \frac{J_{LR} A_0^2}{a}  d\ell +
  J_{LR} A_0^2 a (\partial_x \theta)^2 d\ell +\ldots
  \right],    
\end{eqnarray}
so that
\begin{eqnarray}
  \frac{d}{d\ell} \left(\frac{uK}{2\pi}\right) =  J_{LR} A_0^2
   a,  \nonumber \\
  \frac{d}{d\ell} \left(\frac{K}{2\pi u}\right) = 0.
\end{eqnarray}
By combining these two equations, we arrive at
(\ref{eq:u-rg})--~(\ref{eq:K-rg}). By rescaling the contribution from
$|x-y|\ge a e^{d\ell}$, to restore a cutoff $a$ and taking into
account the scaling dimension $(4K)^{-1}$ of the operator $\cos
\theta$, we recover~(\ref{eq:jlr-rg}).

\section{Integration of the renormalization group equations}\label{app:integration-rg}

The renormalization group equations~(\ref{eq:rg-reduced}) possess the invariant Eq.~(\ref{eq:rg-invariant}). An equivalent form is
\begin{equation}
  \frac{2 g_{LR}}{3-\alpha} -\left(K -\frac 1 {6-2\alpha}\right)^2 =\mathcal{C'}.  
\end{equation}
Above the separatrix in Fig.~\ref{fig:rgflow}, $\mathcal{C'}>0$, we write
\begin{eqnarray}
  g_{LR}&=&\frac{3-\alpha}{2} \mathcal{C'} \cosh^2 \theta, \\
  K&=&\frac 1 {6-2\alpha} + \sqrt{\mathcal{C'}} \sinh \theta, 
\end{eqnarray}
and we obtain a single differential equation for $\theta$
\begin{equation}
  \frac{d\theta}{d\ell} = \frac{(3-\alpha) \sqrt{\mathcal{C'}} \cosh \theta}{\frac 1 {3-\alpha} + 2 \sqrt{\mathcal{C'}} \sinh \theta},  
\end{equation}
that is integrated in the form
\begin{eqnarray}
&&  \frac{2}{3-\alpha} [\arctan(e^{\theta(\ell)})
   -\arctan(e^{\theta(0)})] + 2  \sqrt{\mathcal{C'}} \ln \frac{\cosh
   \theta(\ell)}{\cosh \theta(0)} \nonumber \\ 
  && = (3-\alpha)  \sqrt{\mathcal{C'}} \ell.  
\end{eqnarray}

For $\mathcal{C'}\ll 1$, the perturbative RG breaks down when $\sqrt{\mathcal{C'}} e^{\theta(\ell^*)} = O(1)$. The leading behavior is then $\ell^* \sim \frac{\pi}{(3-\alpha)^2 \sqrt{\mathcal{C'}}}$, and the correlation length is diverging in a similar manner as in the Kosterlitz-Thouless transition.
Below the separatrix in Fig.~\ref{fig:rgflow}, $\mathcal{C'}<0$, we take
  \begin{eqnarray}
  g_{LR}&=&\frac{3-\alpha}{2} \mathcal{-C'} \sinh^2 \theta, \\
  K&=&\frac 1 {6-2\alpha} + r \sqrt{\mathcal{-C'}} \cosh \theta, 
  \end{eqnarray}
  with $r=\pm 1$. The differential equation becomes
  \begin{equation}
     \frac{d\theta}{d\ell} = \frac{(3-\alpha) \sqrt{\mathcal{-C'}} \sinh \theta}{\frac r {3-\alpha} + 2 \sqrt{\mathcal{-C'}} \cosh \theta},   
\end{equation}

which is integrated in the form
\begin{eqnarray}
   \frac{2r}{3-\alpha} \ln \frac{\tanh
  \theta(\ell)/2}{\tanh\theta(0)/2} + 2  \sqrt{\mathcal{-C'}}  \ln
  \frac{\sinh \theta(\ell)}{\sinh\theta(0)} = (3-\alpha)
  \sqrt{\mathcal{-C'}} \ell.\nonumber \\     
\end{eqnarray}

In particular, for $r=-1$, we find that $\theta(\ell \to +\infty) \to 0$ and at the fixed point, $K^* = 1/(6-2\alpha) -\sqrt{\mathcal{-C'}} $.
On the separatrix, we can directly use Eq.~(\ref{eq:rg-integrated}) to derive
\begin{eqnarray}
&&  \ln \frac{ (6-2\alpha) K(\ell) -1}{(6-2\alpha) K(0) -1} + \frac 1
   {(6-2\alpha) K(0) -1} -\frac{1}{(6-2\alpha) K(\ell) -1} \nonumber
  \\
&&  = \frac{3-\alpha} 2 \ell 
\end{eqnarray}
For $K(0) < 1/(6-2\alpha)$ I get
\begin{eqnarray}
  K(\ell) = \frac 1{2(3-\alpha)} -\frac{1}{(3-\alpha)^2 \ell} + -\frac{2 \ln \ell }{(3-\alpha)^2 \ell^2}+ O(\ell^{-2}).  
\end{eqnarray}
Such expression gives rise to
\begin{equation}
  \int^{\ln(r/a)} \frac{d\ell}{2K(\ell)} = (3-\alpha) \ln(r/a) + 2 \ln[\ln(r/a)] + O(1) 
\end{equation}
and results in
\begin{eqnarray}
  \langle e^{i\theta(r)} e^{-i\theta(0)}\rangle
  &=&\exp\left[-\int^{\ln(r/a)} \frac{d\ell}{2K(\ell)}
      \right] \nonumber \\ 
  &=&\left(\frac a r\right)^{3-\alpha} (\ln (r/a))^{-2},
\end{eqnarray}
giving a logarithmic correction to the fixed point correlation function at the transition between the TLL and the CSB. 

\section{Summation of the series giving the Tomonaga-Luttinger
  exponent at finite temperature}\label{app:summation}

To obtain the TL exponent at low temperature, following
Eq.~(\ref{eq:K-temp}), we need the sum
\begin{equation}
  S=\sum_{l=2}^{+\infty} l^{2-\alpha} \left(\frac{\pi T
      a}{\sinh \frac{\pi T la}{\tilde{u}}}\right)^{\frac 1
    {2\tilde{K}}}.   
\end{equation}
Writing the hyperbolic sine in terms of exponentials, and applying a
Taylor expansion, we find
\begin{eqnarray}
  S &=& \left(\frac{2 \pi T a}{\tilde{u}}\right)^{\frac 1
    {2\tilde{K}}} \sum_{l=2}^{+\infty} \sum_{k=0}^{+\infty}
  l^{2-\alpha} \frac{\Gamma\left(k+\frac 1
  {2\tilde{K}}\right)}{\Gamma\left(\frac 1
  {2\tilde{K}}\right)\Gamma\left(k+1\right)} e^{-\frac{\pi T
  l a}{\tilde{u}} \left(2 k+\frac 1
  {2\tilde{K}}\right)}, \nonumber \\  
  &=& \left(\frac{2 \pi T a}{\tilde{u}}\right)^{\frac 1
    {2\tilde{K}}} \sum_{k=0}^{+\infty} \frac{\Gamma\left(k+\frac 1
  {2\tilde{K}}\right)}{\Gamma\left(\frac 1
  {2\tilde{K}}\right)\Gamma\left(k+1\right)}
      \left[\mathrm{Li}_{\alpha-2} \left(e^{-\frac{\pi T
   a}{\tilde{u}} \left(2 k+\frac 1
      {2\tilde{K}}\right)} \right) \right. \nonumber \\
    && \left.- e^{-\frac{\pi T
   a}{\tilde{u}} \left(2 k+\frac 1
  {2\tilde{K}}\right)} \right].  
\end{eqnarray}
when $\tilde{K} \to +\infty$, only the term $k=0$ can give rise to a
divergent contribution. Using Eq.~(25.12.12) in \cite{olver2010nist},
we find the leading divergence
\begin{equation}
  S \sim \Gamma(3-\alpha) \left(\frac{\pi T a}{2\tilde{u} \tilde{K}}
  \right)^{\alpha -3}  \left(\frac{2 \pi T a}{\tilde{u}}\right)^{\frac 1
    {2\tilde{K}}}.  
\end{equation}
When $\tilde{K} \to +\infty$, we can neglect the last factor. 

\begin{thebibliography}{87}%
\makeatletter
\providecommand \@ifxundefined [1]{%
 \@ifx{#1\undefined}
}%
\providecommand \@ifnum [1]{%
 \ifnum #1\expandafter \@firstoftwo
 \else \expandafter \@secondoftwo
 \fi
}%
\providecommand \@ifx [1]{%
 \ifx #1\expandafter \@firstoftwo
 \else \expandafter \@secondoftwo
 \fi
}%
\providecommand \natexlab [1]{#1}%
\providecommand \enquote  [1]{``#1''}%
\providecommand \bibnamefont  [1]{#1}%
\providecommand \bibfnamefont [1]{#1}%
\providecommand \citenamefont [1]{#1}%
\providecommand \href@noop [0]{\@secondoftwo}%
\providecommand \href [0]{\begingroup \@sanitize@url \@href}%
\providecommand \@href[1]{\@@startlink{#1}\@@href}%
\providecommand \@@href[1]{\endgroup#1\@@endlink}%
\providecommand \@sanitize@url [0]{\catcode `\\12\catcode `\$12\catcode
  `\&12\catcode `\#12\catcode `\^12\catcode `\_12\catcode `\%12\relax}%
\providecommand \@@startlink[1]{}%
\providecommand \@@endlink[0]{}%
\providecommand \url  [0]{\begingroup\@sanitize@url \@url }%
\providecommand \@url [1]{\endgroup\@href {#1}{\urlprefix }}%
\providecommand \urlprefix  [0]{URL }%
\providecommand \Eprint [0]{\href }%
\providecommand \doibase [0]{https://doi.org/}%
\providecommand \selectlanguage [0]{\@gobble}%
\providecommand \bibinfo  [0]{\@secondoftwo}%
\providecommand \bibfield  [0]{\@secondoftwo}%
\providecommand \translation [1]{[#1]}%
\providecommand \BibitemOpen [0]{}%
\providecommand \bibitemStop [0]{}%
\providecommand \bibitemNoStop [0]{.\EOS\space}%
\providecommand \EOS [0]{\spacefactor3000\relax}%
\providecommand \BibitemShut  [1]{\csname bibitem#1\endcsname}%
\let\auto@bib@innerbib\@empty
\bibitem [{\citenamefont {Lewenstein~{\it et
  al.}}(2007)}]{AdvPhys07-Lewenstein}%
  \BibitemOpen
  \bibfield  {author} {\bibinfo {author} {\bibfnamefont {M.}~\bibnamefont
  {Lewenstein~{\it et al.}}},\ }\bibfield  {title} {\bibinfo {title} {Ultracold
  atomic gases in optical lattices: mimicking condensed matter physics and
  beyond},\ }\href {https://doi.org/10.1080/00018730701223200} {\bibfield
  {journal} {\bibinfo  {journal} {Adv. Phys.}\ }\textbf {\bibinfo {volume}
  {56}},\ \bibinfo {pages} {243} (\bibinfo {year} {2007})}\BibitemShut
  {NoStop}%
\bibitem [{\citenamefont {Bloch}\ \emph {et~al.}(2008)\citenamefont {Bloch},
  \citenamefont {Dalibard},\ and\ \citenamefont {Zwerger}}]{Bloch2008}%
  \BibitemOpen
  \bibfield  {author} {\bibinfo {author} {\bibfnamefont {I.}~\bibnamefont
  {Bloch}}, \bibinfo {author} {\bibfnamefont {J.}~\bibnamefont {Dalibard}},\
  and\ \bibinfo {author} {\bibfnamefont {W.}~\bibnamefont {Zwerger}},\
  }\bibfield  {title} {\bibinfo {title} {{Many-body physics with ultracold
  gases}},\ }\href {https://doi.org/10.1103/RevModPhys.80.885} {\bibfield
  {journal} {\bibinfo  {journal} {Reviews of Modern Physics}\ }\textbf
  {\bibinfo {volume} {80}},\ \bibinfo {pages} {885} (\bibinfo {year}
  {2008})}\BibitemShut {NoStop}%
\bibitem [{\citenamefont {Chanda}\ \emph {et~al.}(2025)\citenamefont {Chanda},
  \citenamefont {Barbiero}, \citenamefont {Lewenstein}, \citenamefont {Mark},\
  and\ \citenamefont {Zakrzewski}}]{chanda_2025}%
  \BibitemOpen
  \bibfield  {author} {\bibinfo {author} {\bibfnamefont {T.}~\bibnamefont
  {Chanda}}, \bibinfo {author} {\bibfnamefont {L.}~\bibnamefont {Barbiero}},
  \bibinfo {author} {\bibfnamefont {M.}~\bibnamefont {Lewenstein}}, \bibinfo
  {author} {\bibfnamefont {M.~J.}\ \bibnamefont {Mark}},\ and\ \bibinfo
  {author} {\bibfnamefont {J.}~\bibnamefont {Zakrzewski}},\ }\bibfield  {title}
  {\bibinfo {title} {Recent progress on quantum simulations of non-standard
  {Bose}–{Hubbard} models},\ }\href
  {https://doi.org/10.1088/1361-6633/adc3a7} {\bibfield  {journal} {\bibinfo
  {journal} {Reports on Progress in Physics}\ }\textbf {\bibinfo {volume}
  {88}},\ \bibinfo {pages} {044501} (\bibinfo {year} {2025})}\BibitemShut
  {NoStop}%
\bibitem [{\citenamefont {Johanning}\ \emph {et~al.}(2009)\citenamefont
  {Johanning}, \citenamefont {Varon},\ and\ \citenamefont
  {Wunderlich}}]{johanning_2009}%
  \BibitemOpen
  \bibfield  {author} {\bibinfo {author} {\bibfnamefont {M.}~\bibnamefont
  {Johanning}}, \bibinfo {author} {\bibfnamefont {A.}~\bibnamefont {Varon}},\
  and\ \bibinfo {author} {\bibfnamefont {C.}~\bibnamefont {Wunderlich}},\
  }\bibfield  {title} {\bibinfo {title} {Quantum {Simulations} with {Cold}
  {Trapped} {Ions}},\ }\href {https://doi.org/10.1088/0953-4075/42/15/154009}
  {\bibfield  {journal} {\bibinfo  {journal} {Journal of Physics B: Atomic,
  Molecular and Optical Physics}\ }\textbf {\bibinfo {volume} {42}},\ \bibinfo
  {pages} {154009} (\bibinfo {year} {2009})},\ \bibinfo {note} {arXiv:0905.0118
  [quant-ph]}\BibitemShut {NoStop}%
\bibitem [{\citenamefont {Foss-Feig}\ \emph {et~al.}(2025)\citenamefont
  {Foss-Feig}, \citenamefont {Pagano}, \citenamefont {Potter},\ and\
  \citenamefont {Yao}}]{foss-feig_2025}%
  \BibitemOpen
  \bibfield  {author} {\bibinfo {author} {\bibfnamefont {M.}~\bibnamefont
  {Foss-Feig}}, \bibinfo {author} {\bibfnamefont {G.}~\bibnamefont {Pagano}},
  \bibinfo {author} {\bibfnamefont {A.~C.}\ \bibnamefont {Potter}},\ and\
  \bibinfo {author} {\bibfnamefont {N.~Y.}\ \bibnamefont {Yao}},\ }\bibfield
  {title} {\bibinfo {title} {Progress in {Trapped}-{Ion} {Quantum}
  {Simulation}},\ }\href
  {https://doi.org/10.1146/annurev-conmatphys-032822-045619} {\bibfield
  {journal} {\bibinfo  {journal} {Annual Review of Condensed Matter Physics}\
  }\textbf {\bibinfo {volume} {16}},\ \bibinfo {pages} {172} (\bibinfo {year}
  {2025})},\ \bibinfo {note} {arXiv:2409.02990 [cond-mat,
  physics:quant-ph]}\BibitemShut {NoStop}%
\bibitem [{\citenamefont {Feng}\ \emph {et~al.}(2023)\citenamefont {Feng},
  \citenamefont {Katz}, \citenamefont {Haack}, \citenamefont {Maghrebi},
  \citenamefont {Gorshkov}, \citenamefont {Gong}, \citenamefont {Cetina},\ and\
  \citenamefont {Monroe}}]{feng2023}%
  \BibitemOpen
  \bibfield  {author} {\bibinfo {author} {\bibfnamefont {L.}~\bibnamefont
  {Feng}}, \bibinfo {author} {\bibfnamefont {O.}~\bibnamefont {Katz}}, \bibinfo
  {author} {\bibfnamefont {C.}~\bibnamefont {Haack}}, \bibinfo {author}
  {\bibfnamefont {M.}~\bibnamefont {Maghrebi}}, \bibinfo {author}
  {\bibfnamefont {A.~V.}\ \bibnamefont {Gorshkov}}, \bibinfo {author}
  {\bibfnamefont {Z.}~\bibnamefont {Gong}}, \bibinfo {author} {\bibfnamefont
  {M.}~\bibnamefont {Cetina}},\ and\ \bibinfo {author} {\bibfnamefont
  {C.}~\bibnamefont {Monroe}},\ }\bibfield  {title} {\bibinfo {title}
  {Continuous symmetry breaking in a trapped-ion spin chain},\ }\href
  {https://doi.org/10.1038/s41586-023-06656-7} {\bibfield  {journal} {\bibinfo
  {journal} {Nature}\ }\textbf {\bibinfo {volume} {623}},\ \bibinfo {pages}
  {713} (\bibinfo {year} {2023})}\BibitemShut {NoStop}%
\bibitem [{\citenamefont {Emperauger}\ \emph {et~al.}(2025)\citenamefont
  {Emperauger}, \citenamefont {Qiao}, \citenamefont {Chen}, \citenamefont
  {Caleca}, \citenamefont {Bocini}, \citenamefont {Bintz}, \citenamefont
  {Bornet}, \citenamefont {Martin}, \citenamefont {Gély}, \citenamefont
  {Klein}, \citenamefont {Barredo}, \citenamefont {Chatterjee}, \citenamefont
  {Yao}, \citenamefont {Mezzacapo}, \citenamefont {Lahaye}, \citenamefont
  {Roscilde},\ and\ \citenamefont {Browaeys}}]{emperauger2025}%
  \BibitemOpen
  \bibfield  {author} {\bibinfo {author} {\bibfnamefont {G.}~\bibnamefont
  {Emperauger}}, \bibinfo {author} {\bibfnamefont {M.}~\bibnamefont {Qiao}},
  \bibinfo {author} {\bibfnamefont {C.}~\bibnamefont {Chen}}, \bibinfo {author}
  {\bibfnamefont {F.}~\bibnamefont {Caleca}}, \bibinfo {author} {\bibfnamefont
  {S.}~\bibnamefont {Bocini}}, \bibinfo {author} {\bibfnamefont
  {M.}~\bibnamefont {Bintz}}, \bibinfo {author} {\bibfnamefont
  {G.}~\bibnamefont {Bornet}}, \bibinfo {author} {\bibfnamefont
  {R.}~\bibnamefont {Martin}}, \bibinfo {author} {\bibfnamefont
  {B.}~\bibnamefont {Gély}}, \bibinfo {author} {\bibfnamefont
  {L.}~\bibnamefont {Klein}}, \bibinfo {author} {\bibfnamefont
  {D.}~\bibnamefont {Barredo}}, \bibinfo {author} {\bibfnamefont
  {S.}~\bibnamefont {Chatterjee}}, \bibinfo {author} {\bibfnamefont
  {N.}~\bibnamefont {Yao}}, \bibinfo {author} {\bibfnamefont {F.}~\bibnamefont
  {Mezzacapo}}, \bibinfo {author} {\bibfnamefont {T.}~\bibnamefont {Lahaye}},
  \bibinfo {author} {\bibfnamefont {T.}~\bibnamefont {Roscilde}},\ and\
  \bibinfo {author} {\bibfnamefont {A.}~\bibnamefont {Browaeys}},\ }\href
  {http://arxiv.org/abs/2501.08179} {\bibinfo {title} {Tomonaga-{Luttinger}
  {Liquid} {Behavior} in a {Rydberg}-encoded {Spin} {Chain}}} (\bibinfo {year}
  {2025}),\ \bibinfo {note} {arXiv:2501.08179 [quant-ph]}\BibitemShut {NoStop}%
\bibitem [{\citenamefont {Giergiel}\ \emph {et~al.}(2025)\citenamefont
  {Giergiel}, \citenamefont {Hannaford},\ and\ \citenamefont
  {Sacha}}]{giergiel_2025}%
  \BibitemOpen
  \bibfield  {author} {\bibinfo {author} {\bibfnamefont {K.}~\bibnamefont
  {Giergiel}}, \bibinfo {author} {\bibfnamefont {P.}~\bibnamefont
  {Hannaford}},\ and\ \bibinfo {author} {\bibfnamefont {K.}~\bibnamefont
  {Sacha}},\ }\href {https://doi.org/10.48550/arXiv.2406.06387} {\bibinfo
  {title} {Time-tronics: from temporal printed circuit board to quantum
  computer}} (\bibinfo {year} {2025}),\ \bibinfo {note} {arXiv:2406.06387
  [cond-mat]}\BibitemShut {NoStop}%
\bibitem [{\citenamefont {Yang}\ \emph {et~al.}(2024)\citenamefont {Yang},
  \citenamefont {Schumm},\ and\ \citenamefont {Sandvik}}]{yang2024}%
  \BibitemOpen
  \bibfield  {author} {\bibinfo {author} {\bibfnamefont {S.}~\bibnamefont
  {Yang}}, \bibinfo {author} {\bibfnamefont {G.}~\bibnamefont {Schumm}},\ and\
  \bibinfo {author} {\bibfnamefont {A.~W.}\ \bibnamefont {Sandvik}},\ }\href
  {http://arxiv.org/abs/2412.15168} {\bibinfo {title} {Dynamic structure factor
  of a spin-1/2 {Heisenberg} chain with long-range interactions}} (\bibinfo
  {year} {2024}),\ \bibinfo {note} {arXiv:2412.15168 [cond-mat]}\BibitemShut
  {NoStop}%
\bibitem [{\citenamefont {Tung}\ and\ \citenamefont {Guo}(2011)}]{tung2011}%
  \BibitemOpen
  \bibfield  {author} {\bibinfo {author} {\bibfnamefont {J.~C.}\ \bibnamefont
  {Tung}}\ and\ \bibinfo {author} {\bibfnamefont {G.~Y.}\ \bibnamefont {Guo}},\
  }\bibfield  {title} {\bibinfo {title} {Ab initio studies of spin-spiral waves
  and exchange interactions in $3d$ transition metal atomic chains},\ }\href
  {https://doi.org/10.1103/PhysRevB.83.144403} {\bibfield  {journal} {\bibinfo
  {journal} {Phys. Rev. B}\ }\textbf {\bibinfo {volume} {83}},\ \bibinfo
  {pages} {144403} (\bibinfo {year} {2011})}\BibitemShut {NoStop}%
\bibitem [{\citenamefont {Landau}\ and\ \citenamefont
  {Lifshitz}(1986)}]{landau-statmech-english}%
  \BibitemOpen
  \bibfield  {author} {\bibinfo {author} {\bibfnamefont {L.~D.}\ \bibnamefont
  {Landau}}\ and\ \bibinfo {author} {\bibfnamefont {I.~M.}\ \bibnamefont
  {Lifshitz}},\ }\href@noop {} {\emph {\bibinfo {title} {Statistical Physics.
  3rd edition}}}\ (\bibinfo  {publisher} {Pergamon Press},\ \bibinfo {address}
  {Oxford},\ \bibinfo {year} {1986})\BibitemShut {NoStop}%
\bibitem [{\citenamefont {Hohenberg}(1967)}]{hohenberg67_theorem}%
  \BibitemOpen
  \bibfield  {author} {\bibinfo {author} {\bibfnamefont {P.~C.}\ \bibnamefont
  {Hohenberg}},\ }\bibfield  {title} {\bibinfo {title} {Existence of long-range
  order in one and two dimensions},\ }\href@noop {} {\bibfield  {journal}
  {\bibinfo  {journal} {Physical Review}\ }\textbf {\bibinfo {volume} {158}},\
  \bibinfo {pages} {383} (\bibinfo {year} {1967})}\BibitemShut {NoStop}%
\bibitem [{\citenamefont {Mermin}\ and\ \citenamefont
  {Wagner}(1966)}]{mermin_wagner_theorem}%
  \BibitemOpen
  \bibfield  {author} {\bibinfo {author} {\bibfnamefont {N.~D.}\ \bibnamefont
  {Mermin}}\ and\ \bibinfo {author} {\bibfnamefont {H.}~\bibnamefont
  {Wagner}},\ }\bibfield  {title} {\bibinfo {title} {Absence of ferromagnetism
  or antiferromagnetism in one- or two-dimensional isotropic heisenberg
  models},\ }\href {https://doi.org/10.1103/PhysRevLett.17.1133} {\bibfield
  {journal} {\bibinfo  {journal} {Phys. Rev. Lett.}\ }\textbf {\bibinfo
  {volume} {17}},\ \bibinfo {pages} {1133} (\bibinfo {year}
  {1966})}\BibitemShut {NoStop}%
\bibitem [{\citenamefont {Mermin}(1968)}]{mermin_theorem}%
  \BibitemOpen
  \bibfield  {author} {\bibinfo {author} {\bibfnamefont {N.~D.}\ \bibnamefont
  {Mermin}},\ }\href@noop {} {\bibfield  {journal} {\bibinfo  {journal}
  {Physical Review}\ }\textbf {\bibinfo {volume} {176}},\ \bibinfo {pages}
  {250} (\bibinfo {year} {1968})}\BibitemShut {NoStop}%
\bibitem [{\citenamefont {Coleman}(1973)}]{coleman1973}%
  \BibitemOpen
  \bibfield  {author} {\bibinfo {author} {\bibfnamefont {S.}~\bibnamefont
  {Coleman}},\ }\bibfield  {title} {\bibinfo {title} {There are no goldstone
  bosons in two dimensions},\ }\href
  {http://projecteuclid.org/euclid.cmp/1103859034} {\bibfield  {journal}
  {\bibinfo  {journal} {Comm. Math. Phys.}\ }\textbf {\bibinfo {volume} {31}},\
  \bibinfo {pages} {259} (\bibinfo {year} {1973})}\BibitemShut {NoStop}%
\bibitem [{\citenamefont {{Pitaevskii}}\ and\ \citenamefont
  {{Stringari}}(1991)}]{pitaevskii1991}%
  \BibitemOpen
  \bibfield  {author} {\bibinfo {author} {\bibfnamefont {L.}~\bibnamefont
  {{Pitaevskii}}}\ and\ \bibinfo {author} {\bibfnamefont {S.}~\bibnamefont
  {{Stringari}}},\ }\bibfield  {title} {\bibinfo {title} {{Uncertainty
  principle, quantum fluctuations, and broken symmetries}},\ }\href
  {https://doi.org/10.1007/BF00682193} {\bibfield  {journal} {\bibinfo
  {journal} {Journal of Low Temperature Physics}\ }\textbf {\bibinfo {volume}
  {85}},\ \bibinfo {pages} {377} (\bibinfo {year} {1991})}\BibitemShut
  {NoStop}%
\bibitem [{\citenamefont {Gelfert}\ and\ \citenamefont
  {Nolting}(2001)}]{gelfert2001}%
  \BibitemOpen
  \bibfield  {author} {\bibinfo {author} {\bibfnamefont {A.}~\bibnamefont
  {Gelfert}}\ and\ \bibinfo {author} {\bibfnamefont {W.}~\bibnamefont
  {Nolting}},\ }\bibfield  {title} {\bibinfo {title} {The absence of
  finite-temperature phase transitions in low-dimensional many-body models: a
  survey and new results},\ }\href
  {https://doi.org/10.1088/0953-8984/13/27/201} {\bibfield  {journal} {\bibinfo
   {journal} {Journal of Physics: Condensed Matter}\ }\textbf {\bibinfo
  {volume} {13}},\ \bibinfo {pages} {R505} (\bibinfo {year} {2001})},\ \bibinfo
  {note} {arXiv: cond-mat/0106090}\BibitemShut {NoStop}%
\bibitem [{\citenamefont {Giamarchi}(2004)}]{giamarchi_book_1d}%
  \BibitemOpen
  \bibfield  {author} {\bibinfo {author} {\bibfnamefont {T.}~\bibnamefont
  {Giamarchi}},\ }\href@noop {} {\emph {\bibinfo {title} {Quantum Physics in
  One Dimension}}},\ \bibinfo {series} {International series of monographs on
  physics}, Vol.\ \bibinfo {volume} {121}\ (\bibinfo  {publisher} {Oxford
  University Press},\ \bibinfo {address} {Oxford},\ \bibinfo {year}
  {2004})\BibitemShut {NoStop}%
\bibitem [{\citenamefont {Tomonaga}(1950)}]{tomonaga50_1D_electron_gas}%
  \BibitemOpen
  \bibfield  {author} {\bibinfo {author} {\bibfnamefont {S.}~\bibnamefont
  {Tomonaga}},\ }\bibfield  {title} {\bibinfo {title} {Remark on bloch's method
  of sound waves applied to many-fermion problesm},\ }\href@noop {} {\bibfield
  {journal} {\bibinfo  {journal} {Prog. Theor. Phys.}\ }\textbf {\bibinfo
  {volume} {5}},\ \bibinfo {pages} {544} (\bibinfo {year} {1950})}\BibitemShut
  {NoStop}%
\bibitem [{\citenamefont {Luttinger}(1963)}]{luttinger_model}%
  \BibitemOpen
  \bibfield  {author} {\bibinfo {author} {\bibfnamefont {J.~M.}\ \bibnamefont
  {Luttinger}},\ }\bibfield  {title} {\bibinfo {title} {An exact soluble model
  of a many-fermion system},\ }\href@noop {} {\bibfield  {journal} {\bibinfo
  {journal} {Journal of Mathematical Physics}\ }\textbf {\bibinfo {volume}
  {4}},\ \bibinfo {pages} {1154} (\bibinfo {year} {1963})}\BibitemShut
  {NoStop}%
\bibitem [{\citenamefont
  {Haldane}(1981{\natexlab{a}})}]{haldane_effective_harmonic_fluid_approach}%
  \BibitemOpen
  \bibfield  {author} {\bibinfo {author} {\bibfnamefont {F.~D.~M.}\
  \bibnamefont {Haldane}},\ }\bibfield  {title} {\bibinfo {title} {Effective
  harmonic-fluid approach to low-energy properties of one-dimensional quantum
  fluids},\ }\href@noop {} {\bibfield  {journal} {\bibinfo  {journal} {Physical
  Review Letters}\ }\textbf {\bibinfo {volume} {47}},\ \bibinfo {pages} {1840}
  (\bibinfo {year} {1981}{\natexlab{a}})}\BibitemShut {NoStop}%
\bibitem [{\citenamefont {Lecheminant}(2005)}]{lecheminant_revue_1d}%
  \BibitemOpen
  \bibfield  {author} {\bibinfo {author} {\bibfnamefont {P.}~\bibnamefont
  {Lecheminant}},\ }\bibfield  {title} {\bibinfo {title} {One-dimensional
  quantum spin liquids},\ }in\ \href@noop {} {\emph {\bibinfo {booktitle}
  {Frustrated spin systems}}},\ \bibinfo {editor} {edited by\ \bibinfo {editor}
  {\bibfnamefont {H.~T.}\ \bibnamefont {Diep}}}\ (\bibinfo  {publisher} {World
  Scientific},\ \bibinfo {address} {Singapore},\ \bibinfo {year} {2005})\
  \bibinfo {note} {and references therein}\BibitemShut {NoStop}%
\bibitem [{\citenamefont {Dyson}(1969)}]{dyson1969}%
  \BibitemOpen
  \bibfield  {author} {\bibinfo {author} {\bibfnamefont {F.~J.}\ \bibnamefont
  {Dyson}},\ }\bibfield  {title} {\bibinfo {title} {Non-existence of
  spontaneous magnetization in a one-dimensional {Ising} ferromagnet},\ }\href
  {https://doi.org/10.1007/BF01661575} {\bibfield  {journal} {\bibinfo
  {journal} {Communications in Mathematical Physics}\ }\textbf {\bibinfo
  {volume} {12}},\ \bibinfo {pages} {212} (\bibinfo {year} {1969})}\BibitemShut
  {NoStop}%
\bibitem [{\citenamefont {Thouless}(1969)}]{thouless1969}%
  \BibitemOpen
  \bibfield  {author} {\bibinfo {author} {\bibfnamefont {D.~J.}\ \bibnamefont
  {Thouless}},\ }\bibfield  {title} {\bibinfo {title} {Long-{Range} {Order} in
  {One}-{Dimensional} {Ising} {Systems}},\ }\href
  {https://doi.org/10.1103/PhysRev.187.732} {\bibfield  {journal} {\bibinfo
  {journal} {Physical Review}\ }\textbf {\bibinfo {volume} {187}},\ \bibinfo
  {pages} {732} (\bibinfo {year} {1969})}\BibitemShut {NoStop}%
\bibitem [{\citenamefont {Fisher}\ \emph {et~al.}(1972)\citenamefont {Fisher},
  \citenamefont {Ma},\ and\ \citenamefont {Nickel}}]{fisher1972}%
  \BibitemOpen
  \bibfield  {author} {\bibinfo {author} {\bibfnamefont {M.~E.}\ \bibnamefont
  {Fisher}}, \bibinfo {author} {\bibfnamefont {S.-k.}\ \bibnamefont {Ma}},\
  and\ \bibinfo {author} {\bibfnamefont {B.~G.}\ \bibnamefont {Nickel}},\
  }\bibfield  {title} {\bibinfo {title} {Critical exponents for long-range
  interactions},\ }\href {https://doi.org/10.1103/PhysRevLett.29.917}
  {\bibfield  {journal} {\bibinfo  {journal} {Phys. Rev. Lett.}\ }\textbf
  {\bibinfo {volume} {29}},\ \bibinfo {pages} {917} (\bibinfo {year}
  {1972})}\BibitemShut {NoStop}%
\bibitem [{\citenamefont {Defenu}\ \emph {et~al.}(2023)\citenamefont {Defenu},
  \citenamefont {Donner}, \citenamefont {Macrì}, \citenamefont {Pagano},
  \citenamefont {Ruffo},\ and\ \citenamefont {Trombettoni}}]{defenu2023}%
  \BibitemOpen
  \bibfield  {author} {\bibinfo {author} {\bibfnamefont {N.}~\bibnamefont
  {Defenu}}, \bibinfo {author} {\bibfnamefont {T.}~\bibnamefont {Donner}},
  \bibinfo {author} {\bibfnamefont {T.}~\bibnamefont {Macrì}}, \bibinfo
  {author} {\bibfnamefont {G.}~\bibnamefont {Pagano}}, \bibinfo {author}
  {\bibfnamefont {S.}~\bibnamefont {Ruffo}},\ and\ \bibinfo {author}
  {\bibfnamefont {A.}~\bibnamefont {Trombettoni}},\ }\bibfield  {title}
  {\bibinfo {title} {Long-range interacting quantum systems},\ }\href
  {https://doi.org/10.1103/RevModPhys.95.035002} {\bibfield  {journal}
  {\bibinfo  {journal} {Reviews of Modern Physics}\ }\textbf {\bibinfo {volume}
  {95}},\ \bibinfo {pages} {035002} (\bibinfo {year} {2023})},\ \bibinfo {note}
  {publisher: American Physical Society}\BibitemShut {NoStop}%
\bibitem [{\citenamefont {Nakano}\ and\ \citenamefont
  {Takahashi}(1995)}]{nakano1995}%
  \BibitemOpen
  \bibfield  {author} {\bibinfo {author} {\bibfnamefont {H.}~\bibnamefont
  {Nakano}}\ and\ \bibinfo {author} {\bibfnamefont {M.}~\bibnamefont
  {Takahashi}},\ }\bibfield  {title} {\bibinfo {title} {Magnetic properties of
  quantum heisenberg ferromagnets with long-range interactions},\ }\href
  {https://doi.org/10.1103/PhysRevB.52.6606} {\bibfield  {journal} {\bibinfo
  {journal} {Phys. Rev. B}\ }\textbf {\bibinfo {volume} {52}},\ \bibinfo
  {pages} {6606} (\bibinfo {year} {1995})}\BibitemShut {NoStop}%
\bibitem [{\citenamefont {Bruno}(2001)}]{bruno2001}%
  \BibitemOpen
  \bibfield  {author} {\bibinfo {author} {\bibfnamefont {P.}~\bibnamefont
  {Bruno}},\ }\bibfield  {title} {\bibinfo {title} {Absence of spontaneous
  magnetic order at nonzero temperature in one- and two-dimensional heisenberg
  and $\mathit{XY}$ systems with long-range interactions},\ }\href
  {https://doi.org/10.1103/PhysRevLett.87.137203} {\bibfield  {journal}
  {\bibinfo  {journal} {Phys. Rev. Lett.}\ }\textbf {\bibinfo {volume} {87}},\
  \bibinfo {pages} {137203} (\bibinfo {year} {2001})}\BibitemShut {NoStop}%
\bibitem [{\citenamefont {Parreira}\ \emph {et~al.}(1997)\citenamefont
  {Parreira}, \citenamefont {Bolina},\ and\ \citenamefont
  {Perez}}]{parreira97_longrange1d_neel}%
  \BibitemOpen
  \bibfield  {author} {\bibinfo {author} {\bibfnamefont {J.~R.}\ \bibnamefont
  {Parreira}}, \bibinfo {author} {\bibfnamefont {O.}~\bibnamefont {Bolina}},\
  and\ \bibinfo {author} {\bibfnamefont {J.~F.}\ \bibnamefont {Perez}},\
  }\bibfield  {title} {\bibinfo {title} {N\'eel order in the ground state of
  heisenberg antiferromagnetic chains with long-range interactions},\
  }\href@noop {} {\bibfield  {journal} {\bibinfo  {journal} {Journal of Physics
  A}\ }\textbf {\bibinfo {volume} {30}},\ \bibinfo {pages} {1095} (\bibinfo
  {year} {1997})}\BibitemShut {NoStop}%
\bibitem [{\citenamefont {Laflorencie}\ \emph {et~al.}(2005)\citenamefont
  {Laflorencie}, \citenamefont {Affleck},\ and\ \citenamefont
  {Berciu}}]{laflorencie2005}%
  \BibitemOpen
  \bibfield  {author} {\bibinfo {author} {\bibfnamefont {N.}~\bibnamefont
  {Laflorencie}}, \bibinfo {author} {\bibfnamefont {I.}~\bibnamefont
  {Affleck}},\ and\ \bibinfo {author} {\bibfnamefont {M.}~\bibnamefont
  {Berciu}},\ }\bibfield  {title} {\bibinfo {title} {Critical phenomena and
  quantum phase transition in long range {Heisenberg} antiferromagnetic
  chains},\ }\href {https://doi.org/10.1088/1742-5468/2005/12/P12001}
  {\bibfield  {journal} {\bibinfo  {journal} {Journal of Statistical Mechanics:
  Theory and Experiment}\ }\textbf {\bibinfo {volume} {2005}},\ \bibinfo
  {pages} {P12001} (\bibinfo {year} {2005})}\BibitemShut {NoStop}%
\bibitem [{\citenamefont {Pires}(1995)}]{pires_easy_1995}%
  \BibitemOpen
  \bibfield  {author} {\bibinfo {author} {\bibfnamefont {A.~S.~T.}\
  \bibnamefont {Pires}},\ }\bibfield  {title} {\bibinfo {title} {Easy plane
  {Heisenberg} model with long-range ferromagnetic interactions},\ }\href
  {https://doi.org/10.1016/0375-9601(95)00309-Q} {\bibfield  {journal}
  {\bibinfo  {journal} {Physics Letters A}\ }\textbf {\bibinfo {volume}
  {202}},\ \bibinfo {pages} {309} (\bibinfo {year} {1995})}\BibitemShut
  {NoStop}%
\bibitem [{\citenamefont {Yusuf}\ \emph {et~al.}(2004)\citenamefont {Yusuf},
  \citenamefont {Joshi},\ and\ \citenamefont {Yang}}]{yusuf2004}%
  \BibitemOpen
  \bibfield  {author} {\bibinfo {author} {\bibfnamefont {E.}~\bibnamefont
  {Yusuf}}, \bibinfo {author} {\bibfnamefont {A.}~\bibnamefont {Joshi}},\ and\
  \bibinfo {author} {\bibfnamefont {K.}~\bibnamefont {Yang}},\ }\bibfield
  {title} {\bibinfo {title} {Spin waves in antiferromagnetic spin chains with
  long-range interactions},\ }\href
  {https://doi.org/10.1103/PhysRevB.69.144412} {\bibfield  {journal} {\bibinfo
  {journal} {Phys. Rev. B}\ }\textbf {\bibinfo {volume} {69}},\ \bibinfo
  {pages} {144412} (\bibinfo {year} {2004})}\BibitemShut {NoStop}%
\bibitem [{\citenamefont {Maghrebi}\ \emph {et~al.}(2017)\citenamefont
  {Maghrebi}, \citenamefont {Gong},\ and\ \citenamefont
  {Gorshkov}}]{maghrebi2017}%
  \BibitemOpen
  \bibfield  {author} {\bibinfo {author} {\bibfnamefont {M.~F.}\ \bibnamefont
  {Maghrebi}}, \bibinfo {author} {\bibfnamefont {Z.-X.}\ \bibnamefont {Gong}},\
  and\ \bibinfo {author} {\bibfnamefont {A.~V.}\ \bibnamefont {Gorshkov}},\
  }\bibfield  {title} {\bibinfo {title} {Continuous {Symmetry} {Breaking} in
  {1D} {Long}-{Range} {Interacting} {Quantum} {Systems}},\ }\href
  {https://doi.org/10.1103/PhysRevLett.119.023001} {\bibfield  {journal}
  {\bibinfo  {journal} {Physical Review Letters}\ }\textbf {\bibinfo {volume}
  {119}},\ \bibinfo {pages} {023001} (\bibinfo {year} {2017})}\BibitemShut
  {NoStop}%
\bibitem [{\citenamefont {Ren}\ \emph {et~al.}(2020)\citenamefont {Ren},
  \citenamefont {You},\ and\ \citenamefont {Wang}}]{ren2020}%
  \BibitemOpen
  \bibfield  {author} {\bibinfo {author} {\bibfnamefont {J.}~\bibnamefont
  {Ren}}, \bibinfo {author} {\bibfnamefont {W.-L.}\ \bibnamefont {You}},\ and\
  \bibinfo {author} {\bibfnamefont {X.}~\bibnamefont {Wang}},\ }\bibfield
  {title} {\bibinfo {title} {Entanglements and correlations of one-dimensional
  quantum spin-1/2 chain with anisotropic power-law long range interactions},\
  }\href {https://doi.org/10.1103/PhysRevB.101.094410} {\bibfield  {journal}
  {\bibinfo  {journal} {Physical Review B}\ }\textbf {\bibinfo {volume}
  {101}},\ \bibinfo {pages} {094410} (\bibinfo {year} {2020})},\ \bibinfo
  {note} {arXiv: 2003.04472}\BibitemShut {NoStop}%
\bibitem [{\citenamefont {Schneider}\ \emph {et~al.}(2022)\citenamefont
  {Schneider}, \citenamefont {Thomson},\ and\ \citenamefont
  {Sanchez-Palencia}}]{schneider2022}%
  \BibitemOpen
  \bibfield  {author} {\bibinfo {author} {\bibfnamefont {J.~T.}\ \bibnamefont
  {Schneider}}, \bibinfo {author} {\bibfnamefont {S.~J.}\ \bibnamefont
  {Thomson}},\ and\ \bibinfo {author} {\bibfnamefont {L.}~\bibnamefont
  {Sanchez-Palencia}},\ }\bibfield  {title} {\bibinfo {title} {Entanglement
  spectrum and quantum phase diagram of the long-range {XXZ} chain},\ }\href
  {https://doi.org/10.1103/PhysRevB.106.014306} {\bibfield  {journal} {\bibinfo
   {journal} {Physical Review B}\ }\textbf {\bibinfo {volume} {106}},\ \bibinfo
  {pages} {014306} (\bibinfo {year} {2022})},\ \bibinfo {note}
  {arXiv:2202.13343 [cond-mat]}\BibitemShut {NoStop}%
\bibitem [{\citenamefont {Haldane}(1988)}]{haldane_inv_square}%
  \BibitemOpen
  \bibfield  {author} {\bibinfo {author} {\bibfnamefont {F.~D.~M.}\
  \bibnamefont {Haldane}},\ }\href@noop {} {\bibfield  {journal} {\bibinfo
  {journal} {Physical Review Letters}\ }\textbf {\bibinfo {volume} {60}},\
  \bibinfo {pages} {635} (\bibinfo {year} {1988})}\BibitemShut {NoStop}%
\bibitem [{\citenamefont {Shastry}(1988)}]{shastry_inv_square}%
  \BibitemOpen
  \bibfield  {author} {\bibinfo {author} {\bibfnamefont {B.~S.}\ \bibnamefont
  {Shastry}},\ }\href@noop {} {\bibfield  {journal} {\bibinfo  {journal}
  {Physical Review Letters}\ }\textbf {\bibinfo {volume} {60}},\ \bibinfo
  {pages} {639} (\bibinfo {year} {1988})}\BibitemShut {NoStop}%
\bibitem [{\citenamefont {Calogero}(1969)}]{calogero69_model2}%
  \BibitemOpen
  \bibfield  {author} {\bibinfo {author} {\bibfnamefont {F.}~\bibnamefont
  {Calogero}},\ }\bibfield  {title} {\bibinfo {title} {Ground state of a
  one‐dimensional n‐body system},\ }\href@noop {} {\bibfield  {journal}
  {\bibinfo  {journal} {Journal of Mathematical Physics}\ }\textbf {\bibinfo
  {volume} {10}},\ \bibinfo {pages} {2197} (\bibinfo {year}
  {1969})}\BibitemShut {NoStop}%
\bibitem [{\citenamefont {Sutherland}(1971)}]{sutherland71_model1}%
  \BibitemOpen
  \bibfield  {author} {\bibinfo {author} {\bibfnamefont {B.}~\bibnamefont
  {Sutherland}},\ }\bibfield  {title} {\bibinfo {title} {Quantum many‐body
  problem in one dimension: Ground state},\ }\href@noop {} {\bibfield
  {journal} {\bibinfo  {journal} {Journal of Mathematical Physics}\ }\textbf
  {\bibinfo {volume} {12}},\ \bibinfo {pages} {246} (\bibinfo {year}
  {1971})}\BibitemShut {NoStop}%
\bibitem [{\citenamefont {Schulz}(1993)}]{schulz_wigner_1d}%
  \BibitemOpen
  \bibfield  {author} {\bibinfo {author} {\bibfnamefont {H.~J.}\ \bibnamefont
  {Schulz}},\ }\href@noop {} {\bibfield  {journal} {\bibinfo  {journal}
  {Physical Review Letters}\ }\textbf {\bibinfo {volume} {71}},\ \bibinfo
  {pages} {1864} (\bibinfo {year} {1993})}\BibitemShut {NoStop}%
\bibitem [{\citenamefont {Inoue}\ and\ \citenamefont
  {Nomura}(2006)}]{inoue_conformal_2006}%
  \BibitemOpen
  \bibfield  {author} {\bibinfo {author} {\bibfnamefont {H.}~\bibnamefont
  {Inoue}}\ and\ \bibinfo {author} {\bibfnamefont {K.}~\bibnamefont {Nomura}},\
  }\bibfield  {title} {\bibinfo {title} {Conformal field theory in the
  {Tomonaga}–{Luttinger} model with the
  1/r\${\textbackslash}upbeta\$long-range interaction},\ }\href
  {https://doi.org/10.1088/0305-4470/39/9/012} {\bibfield  {journal} {\bibinfo
  {journal} {Journal of Physics A: Mathematical and General}\ }\textbf
  {\bibinfo {volume} {39}},\ \bibinfo {pages} {2161} (\bibinfo {year}
  {2006})},\ \bibinfo {note} {publisher: IOP Publishing}\BibitemShut {NoStop}%
\bibitem [{\citenamefont {Casula}\ \emph {et~al.}(2006)\citenamefont {Casula},
  \citenamefont {Sorella},\ and\ \citenamefont
  {Senatore}}]{casula06_coulomb1d_qmc}%
  \BibitemOpen
  \bibfield  {author} {\bibinfo {author} {\bibfnamefont {M.}~\bibnamefont
  {Casula}}, \bibinfo {author} {\bibfnamefont {S.}~\bibnamefont {Sorella}},\
  and\ \bibinfo {author} {\bibfnamefont {G.}~\bibnamefont {Senatore}},\
  }\bibfield  {title} {\bibinfo {title} {Ground state properties of the
  one-dimensional coulomb gas using the lattice regularized diffusion monte
  carlo method},\ }\href@noop {} {\bibfield  {journal} {\bibinfo  {journal}
  {Physical Review B}\ }\textbf {\bibinfo {volume} {74}},\ \bibinfo {pages}
  {245427} (\bibinfo {year} {2006})}\BibitemShut {NoStop}%
\bibitem [{\citenamefont {Coleman}(1975)}]{coleman_equivalence}%
  \BibitemOpen
  \bibfield  {author} {\bibinfo {author} {\bibfnamefont {S.}~\bibnamefont
  {Coleman}},\ }\bibfield  {title} {\bibinfo {title} {Quantum sine-gordon
  equation as the massive thirring model},\ }\href@noop {} {\bibfield
  {journal} {\bibinfo  {journal} {Physical Review D}\ }\textbf {\bibinfo
  {volume} {11}},\ \bibinfo {pages} {2088} (\bibinfo {year}
  {1975})}\BibitemShut {NoStop}%
\bibitem [{\citenamefont {Suzumura}(1979)}]{suzumura_sg}%
  \BibitemOpen
  \bibfield  {author} {\bibinfo {author} {\bibfnamefont {Y.}~\bibnamefont
  {Suzumura}},\ }\bibfield  {title} {\bibinfo {title} {Collective modes and
  response functions for the bgd model},\ }\href@noop {} {\bibfield  {journal}
  {\bibinfo  {journal} {Prog. Theor. Phys.}\ }\textbf {\bibinfo {volume}
  {61}},\ \bibinfo {pages} {1} (\bibinfo {year} {1979})}\BibitemShut {NoStop}%
\bibitem [{\citenamefont {Lewenstein}\ \emph {et~al.}(2007)\citenamefont
  {Lewenstein}, \citenamefont {Sanpera}, \citenamefont {Ahufinger},
  \citenamefont {Damski}, \citenamefont {{Sen De}},\ and\ \citenamefont
  {Sen}}]{lewenstein07_coldatoms_review}%
  \BibitemOpen
  \bibfield  {author} {\bibinfo {author} {\bibfnamefont {M.}~\bibnamefont
  {Lewenstein}}, \bibinfo {author} {\bibfnamefont {A.}~\bibnamefont {Sanpera}},
  \bibinfo {author} {\bibfnamefont {V.}~\bibnamefont {Ahufinger}}, \bibinfo
  {author} {\bibfnamefont {B.}~\bibnamefont {Damski}}, \bibinfo {author}
  {\bibfnamefont {A.}~\bibnamefont {{Sen De}}},\ and\ \bibinfo {author}
  {\bibfnamefont {U.}~\bibnamefont {Sen}},\ }\bibfield  {title} {\bibinfo
  {title} {Ultracold atomic gases in optical lattices: mimicking condensed
  matter physics and beyond},\ }\href@noop {} {\bibfield  {journal} {\bibinfo
  {journal} {Ann. Phys. (NY)}\ }\textbf {\bibinfo {volume} {56}},\ \bibinfo
  {pages} {243} (\bibinfo {year} {2007})},\ \bibinfo {note}
  {cond-mat/0606771}\BibitemShut {NoStop}%
\bibitem [{\citenamefont {Monien}\ \emph {et~al.}(1998)\citenamefont {Monien},
  \citenamefont {Linn},\ and\ \citenamefont {Elstner}}]{monien98_bose_1d}%
  \BibitemOpen
  \bibfield  {author} {\bibinfo {author} {\bibfnamefont {H.}~\bibnamefont
  {Monien}}, \bibinfo {author} {\bibfnamefont {M.}~\bibnamefont {Linn}},\ and\
  \bibinfo {author} {\bibfnamefont {N.}~\bibnamefont {Elstner}},\ }\bibfield
  {title} {\bibinfo {title} {Trapped one-dimensional bose gas as a luttinger
  liquid},\ }\href@noop {} {\bibfield  {journal} {\bibinfo  {journal} {Physical
  Review A}\ }\textbf {\bibinfo {volume} {58}},\ \bibinfo {pages} {R3395}
  (\bibinfo {year} {1998})}\BibitemShut {NoStop}%
\bibitem [{\citenamefont {K{\"u}hner}\ and\ \citenamefont
  {Monien}(1998{\natexlab{a}})}]{kuhner_bose_hubbard_critical_point}%
  \BibitemOpen
  \bibfield  {author} {\bibinfo {author} {\bibfnamefont {T.~D.}\ \bibnamefont
  {K{\"u}hner}}\ and\ \bibinfo {author} {\bibfnamefont {H.}~\bibnamefont
  {Monien}},\ }\bibfield  {title} {\bibinfo {title} {Phases of the
  one-dimensional bose-hubbard model},\ }\href@noop {} {\bibfield  {journal}
  {\bibinfo  {journal} {Physical Review B}\ }\textbf {\bibinfo {volume} {58}},\
  \bibinfo {pages} {14741(R)} (\bibinfo {year}
  {1998}{\natexlab{a}})}\BibitemShut {NoStop}%
\bibitem [{\citenamefont {Matsubara}\ and\ \citenamefont
  {Matsuda}(1956)}]{matsubara1956}%
  \BibitemOpen
  \bibfield  {author} {\bibinfo {author} {\bibfnamefont {T.}~\bibnamefont
  {Matsubara}}\ and\ \bibinfo {author} {\bibfnamefont {H.}~\bibnamefont
  {Matsuda}},\ }\bibfield  {title} {\bibinfo {title} {A lattice model of liquid
  helium, i},\ }\href {https://doi.org/10.1143/PTP.16.569} {\bibfield
  {journal} {\bibinfo  {journal} {Prog. Theor. Phys.}\ }\textbf {\bibinfo
  {volume} {16}},\ \bibinfo {pages} {569} (\bibinfo {year} {1956})}\BibitemShut
  {NoStop}%
\bibitem [{\citenamefont {Fisher}(1967)}]{fisher_xxz}%
  \BibitemOpen
  \bibfield  {author} {\bibinfo {author} {\bibfnamefont {M.~E.}\ \bibnamefont
  {Fisher}},\ }\href@noop {} {\bibfield  {journal} {\bibinfo  {journal} {Rep.
  Prog. Phys.}\ }\textbf {\bibinfo {volume} {30}},\ \bibinfo {pages} {615}
  (\bibinfo {year} {1967})},\ \bibinfo {note} {and references
  therein}\BibitemShut {NoStop}%
\bibitem [{\citenamefont {Orbach}(1958)}]{orbach1959}%
  \BibitemOpen
  \bibfield  {author} {\bibinfo {author} {\bibfnamefont {R.}~\bibnamefont
  {Orbach}},\ }\bibfield  {title} {\bibinfo {title} {Linear antiferromagnetic
  chain with anisotropic coupling},\ }\href
  {https://doi.org/10.1103/PhysRev.112.309} {\bibfield  {journal} {\bibinfo
  {journal} {Phys. Rev.}\ }\textbf {\bibinfo {volume} {112}},\ \bibinfo {pages}
  {309} (\bibinfo {year} {1958})}\BibitemShut {NoStop}%
\bibitem [{\citenamefont {Luther}\ and\ \citenamefont
  {Peschel}(1975)}]{luther_chaine_xxz}%
  \BibitemOpen
  \bibfield  {author} {\bibinfo {author} {\bibfnamefont {A.}~\bibnamefont
  {Luther}}\ and\ \bibinfo {author} {\bibfnamefont {I.}~\bibnamefont
  {Peschel}},\ }\href@noop {} {\bibfield  {journal} {\bibinfo  {journal}
  {Physical Review B}\ }\textbf {\bibinfo {volume} {12}},\ \bibinfo {pages}
  {3908} (\bibinfo {year} {1975})}\BibitemShut {NoStop}%
\bibitem [{\citenamefont {Haldane}(1980)}]{haldane_xxzchain}%
  \BibitemOpen
  \bibfield  {author} {\bibinfo {author} {\bibfnamefont {F.~D.~M.}\
  \bibnamefont {Haldane}},\ }\href@noop {} {\bibfield  {journal} {\bibinfo
  {journal} {Physical Review Letters}\ }\textbf {\bibinfo {volume} {45}},\
  \bibinfo {pages} {1358} (\bibinfo {year} {1980})}\BibitemShut {NoStop}%
\bibitem [{\citenamefont {K{\"u}hner}\ \emph {et~al.}(2000)\citenamefont
  {K{\"u}hner}, \citenamefont {White},\ and\ \citenamefont
  {Monien}}]{kuhner_bosehubbard}%
  \BibitemOpen
  \bibfield  {author} {\bibinfo {author} {\bibfnamefont {T.~D.}\ \bibnamefont
  {K{\"u}hner}}, \bibinfo {author} {\bibfnamefont {S.~R.}\ \bibnamefont
  {White}},\ and\ \bibinfo {author} {\bibfnamefont {H.}~\bibnamefont
  {Monien}},\ }\bibfield  {title} {\bibinfo {title} {{No Title}},\ }\href@noop
  {} {\bibfield  {journal} {\bibinfo  {journal} {Physical Review B}\ }\textbf
  {\bibinfo {volume} {61}},\ \bibinfo {pages} {12474} (\bibinfo {year}
  {2000})}\BibitemShut {NoStop}%
\bibitem [{\citenamefont {Ejima}\ \emph {et~al.}(2011)\citenamefont {Ejima},
  \citenamefont {Fehske},\ and\ \citenamefont {Gebhard}}]{ejima2011}%
  \BibitemOpen
  \bibfield  {author} {\bibinfo {author} {\bibfnamefont {S.}~\bibnamefont
  {Ejima}}, \bibinfo {author} {\bibfnamefont {H.}~\bibnamefont {Fehske}},\ and\
  \bibinfo {author} {\bibfnamefont {F.}~\bibnamefont {Gebhard}},\ }\bibfield
  {title} {\bibinfo {title} {Dynamic properties of the one-dimensional
  {Bose}-{Hubbard} model},\ }\href {https://doi.org/10.1209/0295-5075/93/30002}
  {\bibfield  {journal} {\bibinfo  {journal} {Europhys. Lett.}\ }\textbf
  {\bibinfo {volume} {93}},\ \bibinfo {pages} {30002} (\bibinfo {year}
  {2011})}\BibitemShut {NoStop}%
\bibitem [{\citenamefont {Haldane}(1981{\natexlab{b}})}]{haldane_bosons}%
  \BibitemOpen
  \bibfield  {author} {\bibinfo {author} {\bibfnamefont {F.~D.~M.}\
  \bibnamefont {Haldane}},\ }\href@noop {} {\bibfield  {journal} {\bibinfo
  {journal} {Physical Review Letters}\ }\textbf {\bibinfo {volume} {47}},\
  \bibinfo {pages} {1840} (\bibinfo {year} {1981}{\natexlab{b}})}\BibitemShut
  {NoStop}%
\bibitem [{\citenamefont {Efetov}\ and\ \citenamefont
  {Larkin}(1975)}]{efetov_larkin75}%
  \BibitemOpen
  \bibfield  {author} {\bibinfo {author} {\bibfnamefont {K.~B.}\ \bibnamefont
  {Efetov}}\ and\ \bibinfo {author} {\bibfnamefont {A.~I.}\ \bibnamefont
  {Larkin}},\ }\href@noop {} {\bibfield  {journal} {\bibinfo  {journal} {Sov.
  Phys. JETP}\ }\textbf {\bibinfo {volume} {42}},\ \bibinfo {pages} {390}
  (\bibinfo {year} {1975})}\BibitemShut {NoStop}%
\bibitem [{\citenamefont {K{\"u}hner}\ and\ \citenamefont
  {Monien}(1998{\natexlab{b}})}]{kuhner1998}%
  \BibitemOpen
  \bibfield  {author} {\bibinfo {author} {\bibfnamefont {T.}~\bibnamefont
  {K{\"u}hner}}\ and\ \bibinfo {author} {\bibfnamefont {H.}~\bibnamefont
  {Monien}},\ }\bibfield  {title} {\bibinfo {title} {Phases of the
  one-dimensional {{B}ose}-{Hubbard} model},\ }\href@noop {} {\bibfield
  {journal} {\bibinfo  {journal} {Physical Review B}\ }\textbf {\bibinfo
  {volume} {58}},\ \bibinfo {pages} {R14741} (\bibinfo {year}
  {1998}{\natexlab{b}})}\BibitemShut {NoStop}%
\bibitem [{\citenamefont {Kiely}\ and\ \citenamefont
  {Mueller}(2022)}]{kiely2022}%
  \BibitemOpen
  \bibfield  {author} {\bibinfo {author} {\bibfnamefont {T.~G.}\ \bibnamefont
  {Kiely}}\ and\ \bibinfo {author} {\bibfnamefont {E.~J.}\ \bibnamefont
  {Mueller}},\ }\bibfield  {title} {\bibinfo {title} {Superfluidity in the
  one-dimensional {Bose}-{Hubbard} model},\ }\href
  {https://doi.org/10.1103/PhysRevB.105.134502} {\bibfield  {journal} {\bibinfo
   {journal} {Physical Review B}\ }\textbf {\bibinfo {volume} {105}},\ \bibinfo
  {pages} {134502} (\bibinfo {year} {2022})},\ \bibinfo {note} {arXiv:
  2202.00669}\BibitemShut {NoStop}%
\bibitem [{\citenamefont {Ovchinnikov}(2004)}]{ovchinnikov2004_ff_xx}%
  \BibitemOpen
  \bibfield  {author} {\bibinfo {author} {\bibfnamefont {A.~A.}\ \bibnamefont
  {Ovchinnikov}},\ }\bibfield  {title} {\bibinfo {title} {Formfactors and the
  functional form of correlators in the xx-spin chain},\ }\href@noop {}
  {\bibfield  {journal} {\bibinfo  {journal} {J. Phys.: Condens. Matter}\
  }\textbf {\bibinfo {volume} {16}},\ \bibinfo {pages} {3147} (\bibinfo {year}
  {2004})}\BibitemShut {NoStop}%
\bibitem [{\citenamefont {Lukyanov}\ and\ \citenamefont
  {Terras}(2003)}]{lukyanov_xxz_asymptotics}%
  \BibitemOpen
  \bibfield  {author} {\bibinfo {author} {\bibfnamefont {S.}~\bibnamefont
  {Lukyanov}}\ and\ \bibinfo {author} {\bibfnamefont {V.}~\bibnamefont
  {Terras}},\ }\bibfield  {title} {\bibinfo {title} {Long-distance asymptotics
  of spin-spin correlation functions for the xxz spin chain},\ }\href@noop {}
  {\bibfield  {journal} {\bibinfo  {journal} {Nuclear Physics B}\ }\textbf
  {\bibinfo {volume} {654}},\ \bibinfo {pages} {323} (\bibinfo {year}
  {2003})},\ \bibinfo {note} {hep-th/0206093}\BibitemShut {NoStop}%
\bibitem [{\citenamefont {Hikihara}\ and\ \citenamefont
  {Furusaki}(2004)}]{furusaki_correlations_xxz_magneticfield}%
  \BibitemOpen
  \bibfield  {author} {\bibinfo {author} {\bibfnamefont {T.}~\bibnamefont
  {Hikihara}}\ and\ \bibinfo {author} {\bibfnamefont {A.}~\bibnamefont
  {Furusaki}},\ }\bibfield  {title} {\bibinfo {title} {Correlation amplitudes
  for the spin-1/2 xxz chain in a magnetic field},\ }\href@noop {} {\bibfield
  {journal} {\bibinfo  {journal} {Physical Review B}\ }\textbf {\bibinfo
  {volume} {69}},\ \bibinfo {pages} {06427} (\bibinfo {year}
  {2004})}\BibitemShut {NoStop}%
\bibitem [{\citenamefont {Olver}\ \emph {et~al.}(2010)\citenamefont {Olver},
  \citenamefont {Lozier}, \citenamefont {Boisvert},\ and\ \citenamefont
  {Clark}}]{olver2010nist}%
  \BibitemOpen
  \bibinfo {editor} {\bibfnamefont {F.}~\bibnamefont {Olver}}, \bibinfo
  {editor} {\bibfnamefont {D.}~\bibnamefont {Lozier}}, \bibinfo {editor}
  {\bibfnamefont {R.}~\bibnamefont {Boisvert}},\ and\ \bibinfo {editor}
  {\bibfnamefont {C.}~\bibnamefont {Clark}},\ eds.,\ \href@noop {} {\emph
  {\bibinfo {title} {NIST handbook of mathematical functions}}}\ (\bibinfo
  {publisher} {Cambridge University Press},\ \bibinfo {address} {Cambridge,
  UK},\ \bibinfo {year} {2010})\BibitemShut {NoStop}%
\bibitem [{\citenamefont {Dupuis}(2024)}]{dupuis2024}%
  \BibitemOpen
  \bibfield  {author} {\bibinfo {author} {\bibfnamefont {N.}~\bibnamefont
  {Dupuis}},\ }\bibfield  {title} {\bibinfo {title} {Superfluid--bose-glass
  transition in a system of disordered bosons with long-range hopping in one
  dimension},\ }\href {https://doi.org/10.1103/PhysRevA.110.033315} {\bibfield
  {journal} {\bibinfo  {journal} {Phys. Rev. A}\ }\textbf {\bibinfo {volume}
  {110}},\ \bibinfo {pages} {033315} (\bibinfo {year} {2024})}\BibitemShut
  {NoStop}%
\bibitem [{\citenamefont {Cardy}(1996)}]{cardy_book_renormalization}%
  \BibitemOpen
  \bibfield  {author} {\bibinfo {author} {\bibfnamefont {J.}~\bibnamefont
  {Cardy}},\ }\href@noop {} {\emph {\bibinfo {title} {Scaling and
  Renormalization in Statistical Physics}}}\ (\bibinfo  {publisher} {Cambridge
  University Press},\ \bibinfo {address} {Cambridge},\ \bibinfo {year}
  {1996})\BibitemShut {NoStop}%
\bibitem [{\citenamefont {Giamarchi}\ and\ \citenamefont
  {Schulz}(1987)}]{giamarchi_loc_lettre}%
  \BibitemOpen
  \bibfield  {author} {\bibinfo {author} {\bibfnamefont {T.}~\bibnamefont
  {Giamarchi}}\ and\ \bibinfo {author} {\bibfnamefont {H.~J.}\ \bibnamefont
  {Schulz}},\ }\href@noop {} {\bibfield  {journal} {\bibinfo  {journal}
  {Europhysics Letters}\ }\textbf {\bibinfo {volume} {3}},\ \bibinfo {pages}
  {1287} (\bibinfo {year} {1987})}\BibitemShut {NoStop}%
\bibitem [{\citenamefont {Giamarchi}\ and\ \citenamefont
  {Schulz}(1988)}]{giamarchi_loc}%
  \BibitemOpen
  \bibfield  {author} {\bibinfo {author} {\bibfnamefont {T.}~\bibnamefont
  {Giamarchi}}\ and\ \bibinfo {author} {\bibfnamefont {H.~J.}\ \bibnamefont
  {Schulz}},\ }\bibfield  {title} {\bibinfo {title} {Anderson localization and
  interactions in one-dimensional metals},\ }\href@noop {} {\bibfield
  {journal} {\bibinfo  {journal} {Physical Review B}\ }\textbf {\bibinfo
  {volume} {37}},\ \bibinfo {pages} {325} (\bibinfo {year} {1988})}\BibitemShut
  {NoStop}%
\bibitem [{\citenamefont {Kosterlitz}\ and\ \citenamefont
  {Thouless}(1973)}]{kosterlitz_thouless}%
  \BibitemOpen
  \bibfield  {author} {\bibinfo {author} {\bibfnamefont {J.~M.}\ \bibnamefont
  {Kosterlitz}}\ and\ \bibinfo {author} {\bibfnamefont {D.~J.}\ \bibnamefont
  {Thouless}},\ }\href@noop {} {\bibfield  {journal} {\bibinfo  {journal} {J.
  Phys. C}\ }\textbf {\bibinfo {volume} {6}},\ \bibinfo {pages} {1181}
  (\bibinfo {year} {1973})}\BibitemShut {NoStop}%
\bibitem [{\citenamefont {Kosterlitz}(1974)}]{kosterlitz_renormalisation_xy}%
  \BibitemOpen
  \bibfield  {author} {\bibinfo {author} {\bibfnamefont {J.~M.}\ \bibnamefont
  {Kosterlitz}},\ }\href@noop {} {\bibfield  {journal} {\bibinfo  {journal} {J.
  Phys. C}\ }\textbf {\bibinfo {volume} {7}},\ \bibinfo {pages} {1046}
  (\bibinfo {year} {1974})}\BibitemShut {NoStop}%
\bibitem [{\citenamefont {Orignac}(2004)}]{orignac04_spingap}%
  \BibitemOpen
  \bibfield  {author} {\bibinfo {author} {\bibfnamefont {E.}~\bibnamefont
  {Orignac}},\ }\bibfield  {title} {\bibinfo {title} {Quantitative expression
  of the spin gap via bosonization for a dimerized spin-1/2 chain},\
  }\href@noop {} {\bibfield  {journal} {\bibinfo  {journal} {Eur. Phys. J B}\
  }\textbf {\bibinfo {volume} {39}},\ \bibinfo {pages} {335} (\bibinfo {year}
  {2004})},\ \Eprint {https://arxiv.org/abs/cond-mat/0403175}
  {cond-mat/0403175} \BibitemShut {NoStop}%
\bibitem [{\citenamefont {Watanabe}\ \emph {et~al.}(1993)\citenamefont
  {Watanabe}, \citenamefont {Nomura},\ and\ \citenamefont
  {Takada}}]{watanabe_2ch}%
  \BibitemOpen
  \bibfield  {author} {\bibinfo {author} {\bibfnamefont {H.}~\bibnamefont
  {Watanabe}}, \bibinfo {author} {\bibfnamefont {K.}~\bibnamefont {Nomura}},\
  and\ \bibinfo {author} {\bibfnamefont {S.}~\bibnamefont {Takada}},\
  }\href@noop {} {\bibfield  {journal} {\bibinfo  {journal} {Journal of the
  Physical Society of Japan}\ }\textbf {\bibinfo {volume} {62}},\ \bibinfo
  {pages} {2845} (\bibinfo {year} {1993})}\BibitemShut {NoStop}%
\bibitem [{\citenamefont {Donohue}(2001)}]{donohue_thesis}%
  \BibitemOpen
  \bibfield  {author} {\bibinfo {author} {\bibfnamefont {P.}~\bibnamefont
  {Donohue}},\ }\emph {\bibinfo {title} {Transition de Mott dans les
  Echelles}},\ \href@noop {} {Ph.D. thesis},\ \bibinfo  {school} {Paris XI
  University} (\bibinfo {year} {2001})\BibitemShut {NoStop}%
\bibitem [{\citenamefont {Citro}\ \emph {et~al.}(2004)\citenamefont {Citro},
  \citenamefont {Orignac},\ and\ \citenamefont
  {Giamarchi}}]{citro04_spinpeierls}%
  \BibitemOpen
  \bibfield  {author} {\bibinfo {author} {\bibfnamefont {R.}~\bibnamefont
  {Citro}}, \bibinfo {author} {\bibfnamefont {E.}~\bibnamefont {Orignac}},\
  and\ \bibinfo {author} {\bibfnamefont {T.}~\bibnamefont {Giamarchi}},\
  }\bibfield  {title} {\bibinfo {title} {Adiabatic-antiadiabatic crossover in a
  spin-peierls chain}} (\bibinfo {year} {2004}),\ \bibinfo {note}
  {cond-mat/0411256}\BibitemShut {NoStop}%
\bibitem [{\citenamefont {Cazalilla}\ \emph {et~al.}(2006)\citenamefont
  {Cazalilla}, \citenamefont {Ho},\ and\ \citenamefont
  {Giamarchi}}]{cazalilla_deconfinement_long}%
  \BibitemOpen
  \bibfield  {author} {\bibinfo {author} {\bibfnamefont {M.~A.}\ \bibnamefont
  {Cazalilla}}, \bibinfo {author} {\bibfnamefont {A.~F.}\ \bibnamefont {Ho}},\
  and\ \bibinfo {author} {\bibfnamefont {T.}~\bibnamefont {Giamarchi}},\
  }\href@noop {} {\bibfield  {journal} {\bibinfo  {journal} {New Journal of
  Physics}\ }\textbf {\bibinfo {volume} {8}},\ \bibinfo {pages} {158} (\bibinfo
  {year} {2006})}\BibitemShut {NoStop}%
\bibitem [{\citenamefont {You}\ \emph {et~al.}(2012)\citenamefont {You},
  \citenamefont {Lee}, \citenamefont {Fang}, \citenamefont {Cazalilla},\ and\
  \citenamefont {Wang}}]{you2012}%
  \BibitemOpen
  \bibfield  {author} {\bibinfo {author} {\bibfnamefont {J.-S.}\ \bibnamefont
  {You}}, \bibinfo {author} {\bibfnamefont {H.}~\bibnamefont {Lee}}, \bibinfo
  {author} {\bibfnamefont {S.}~\bibnamefont {Fang}}, \bibinfo {author}
  {\bibfnamefont {M.~A.}\ \bibnamefont {Cazalilla}},\ and\ \bibinfo {author}
  {\bibfnamefont {D.-W.}\ \bibnamefont {Wang}},\ }\bibfield  {title} {\bibinfo
  {title} {Tuning the kosterlitz-thouless transition to zero temperature in
  anisotropic boson systems},\ }\href@noop {} {\bibfield  {journal} {\bibinfo
  {journal} {Phys. Rev. A}\ }\textbf {\bibinfo {volume} {86}},\ \bibinfo
  {pages} {043612} (\bibinfo {year} {2012})}\BibitemShut {NoStop}%
\bibitem [{\citenamefont {Foini}\ and\ \citenamefont
  {Giamarchi}(2015)}]{foini_ladder_quench_cold}%
  \BibitemOpen
  \bibfield  {author} {\bibinfo {author} {\bibfnamefont {L.}~\bibnamefont
  {Foini}}\ and\ \bibinfo {author} {\bibfnamefont {T.}~\bibnamefont
  {Giamarchi}},\ }\bibfield  {title} {\bibinfo {title} {Nonequilibrium dynamics
  of coupled luttinger liquids},\ }\href
  {https://doi.org/10.1103/PhysRevA.91.023627} {\bibfield  {journal} {\bibinfo
  {journal} {Phys. Rev. A}\ }\textbf {\bibinfo {volume} {91}},\ \bibinfo
  {pages} {023627} (\bibinfo {year} {2015})}\BibitemShut {NoStop}%
\bibitem [{\citenamefont {Majumdar}\ \emph {et~al.}(2023)\citenamefont
  {Majumdar}, \citenamefont {Foini}, \citenamefont {Giamarchi},\ and\
  \citenamefont {Rosso}}]{majumdar2023}%
  \BibitemOpen
  \bibfield  {author} {\bibinfo {author} {\bibfnamefont {S.}~\bibnamefont
  {Majumdar}}, \bibinfo {author} {\bibfnamefont {L.}~\bibnamefont {Foini}},
  \bibinfo {author} {\bibfnamefont {T.}~\bibnamefont {Giamarchi}},\ and\
  \bibinfo {author} {\bibfnamefont {A.}~\bibnamefont {Rosso}},\ }\bibfield
  {title} {\bibinfo {title} {Bath-induced phase transition in a {Luttinger}
  liquid},\ }\href {https://doi.org/10.1103/PhysRevB.107.165113} {\bibfield
  {journal} {\bibinfo  {journal} {Physical Review B}\ }\textbf {\bibinfo
  {volume} {107}},\ \bibinfo {pages} {165113} (\bibinfo {year} {2023})},\
  \bibinfo {note} {arXiv:2210.01590 [cond-mat]}\BibitemShut {NoStop}%
\bibitem [{\citenamefont {Villain}(1974)}]{villain1974}%
  \BibitemOpen
  \bibfield  {author} {\bibinfo {author} {\bibfnamefont {J.}~\bibnamefont
  {Villain}},\ }\bibfield  {title} {\bibinfo {title} {{Quantum theory of one-
  and two-dimensional ferro- and antiferromagnets with an easy magnetization
  plane . I. ideal 1-d or 2-d lattices without in-plane anisotropy}},\ }\href
  {https://doi.org/10.1051/jphys:0197400350102700} {\bibfield  {journal}
  {\bibinfo  {journal} {{Journal de Physique}}\ }\textbf {\bibinfo {volume}
  {35}},\ \bibinfo {pages} {27} (\bibinfo {year} {1974})}\BibitemShut {NoStop}%
\bibitem [{\citenamefont {Abrikosov}\ \emph {et~al.}(1963)\citenamefont
  {Abrikosov}, \citenamefont {Gorkov},\ and\ \citenamefont
  {Dzyaloshinski}}]{abrikosov_book}%
  \BibitemOpen
  \bibfield  {author} {\bibinfo {author} {\bibfnamefont {A.~A.}\ \bibnamefont
  {Abrikosov}}, \bibinfo {author} {\bibfnamefont {L.~P.}\ \bibnamefont
  {Gorkov}},\ and\ \bibinfo {author} {\bibfnamefont {I.~E.}\ \bibnamefont
  {Dzyaloshinski}},\ }\href@noop {} {\emph {\bibinfo {title} {Methods of
  Quantum Field Theory in Statistical Physics}}}\ (\bibinfo  {publisher}
  {Dover},\ \bibinfo {address} {New York},\ \bibinfo {year} {1963})\BibitemShut
  {NoStop}%
\bibitem [{\citenamefont {Vanderstraeten}\ \emph {et~al.}(2018)\citenamefont
  {Vanderstraeten}, \citenamefont {Van~Damme}, \citenamefont {Büchler},\ and\
  \citenamefont {Verstraete}}]{vanderstraeten2018}%
  \BibitemOpen
  \bibfield  {author} {\bibinfo {author} {\bibfnamefont {L.}~\bibnamefont
  {Vanderstraeten}}, \bibinfo {author} {\bibfnamefont {M.}~\bibnamefont
  {Van~Damme}}, \bibinfo {author} {\bibfnamefont {H.~P.}\ \bibnamefont
  {Büchler}},\ and\ \bibinfo {author} {\bibfnamefont {F.}~\bibnamefont
  {Verstraete}},\ }\bibfield  {title} {\bibinfo {title} {Quasiparticles in
  {Quantum} {Spin} {Chains} with {Long}-{Range} {Interactions}},\ }\href
  {https://doi.org/10.1103/PhysRevLett.121.090603} {\bibfield  {journal}
  {\bibinfo  {journal} {Physical Review Letters}\ }\textbf {\bibinfo {volume}
  {121}},\ \bibinfo {pages} {090603} (\bibinfo {year} {2018})}\BibitemShut
  {NoStop}%
\bibitem [{\citenamefont {Liu}\ \emph {et~al.}(2024)\citenamefont {Liu},
  \citenamefont {Yi}, \citenamefont {Zhou},\ and\ \citenamefont
  {Zou}}]{liu2024}%
  \BibitemOpen
  \bibfield  {author} {\bibinfo {author} {\bibfnamefont {R.}~\bibnamefont
  {Liu}}, \bibinfo {author} {\bibfnamefont {J.}~\bibnamefont {Yi}}, \bibinfo
  {author} {\bibfnamefont {S.}~\bibnamefont {Zhou}},\ and\ \bibinfo {author}
  {\bibfnamefont {L.}~\bibnamefont {Zou}},\ }\href
  {https://doi.org/10.48550/arXiv.2405.14929} {\bibinfo {title}
  {Lieb-{Schultz}-{Mattis} theorems and generalizations in long-range
  interacting systems}} (\bibinfo {year} {2024}),\ \bibinfo {note}
  {arXiv:2405.14929 [cond-mat]}\BibitemShut {NoStop}%
\bibitem [{\citenamefont {Zhou}\ and\ \citenamefont {Li}(2024)}]{zhou2024}%
  \BibitemOpen
  \bibfield  {author} {\bibinfo {author} {\bibfnamefont {Y.-N.}\ \bibnamefont
  {Zhou}}\ and\ \bibinfo {author} {\bibfnamefont {X.}~\bibnamefont {Li}},\
  }\href {https://doi.org/10.48550/arXiv.2406.08948} {\bibinfo {title}
  {Validity of the {Lieb}-{Schultz}-{Mattis} {Theorem} in {Long}-{Range}
  {Interacting} {Systems}}} (\bibinfo {year} {2024}),\ \bibinfo {note}
  {arXiv:2406.08948 [cond-mat, physics:quant-ph]}\BibitemShut {NoStop}%
\bibitem [{\citenamefont {Ma}(2024)}]{ma2024}%
  \BibitemOpen
  \bibfield  {author} {\bibinfo {author} {\bibfnamefont {R.}~\bibnamefont
  {Ma}},\ }\bibfield  {title} {\bibinfo {title} {Lieb-{Schultz}-{Mattis}
  {Theorem} with {Long}-{Range} {Interactions}},\ }\href
  {https://doi.org/10.48550/arXiv.2405.14949} {\bibfield  {journal} {\bibinfo
  {journal} {Phys. Rev. B}\ }\textbf {\bibinfo {volume} {110}},\ \bibinfo
  {pages} {104412} (\bibinfo {year} {2024})},\ \bibinfo {note}
  {arXiv:2405.14949 [cond-mat]}\BibitemShut {NoStop}%
\bibitem [{\citenamefont {Lieb}\ \emph {et~al.}(1961)\citenamefont {Lieb},
  \citenamefont {Schultz},\ and\ \citenamefont {Mattis}}]{lieb_lsm_theorem}%
  \BibitemOpen
  \bibfield  {author} {\bibinfo {author} {\bibfnamefont {E.}~\bibnamefont
  {Lieb}}, \bibinfo {author} {\bibfnamefont {T.~D.}\ \bibnamefont {Schultz}},\
  and\ \bibinfo {author} {\bibfnamefont {D.}~\bibnamefont {Mattis}},\
  }\href@noop {} {\bibfield  {journal} {\bibinfo  {journal} {Ann. Phys. (NY)}\
  }\textbf {\bibinfo {volume} {16}},\ \bibinfo {pages} {407} (\bibinfo {year}
  {1961})}\BibitemShut {NoStop}%
\bibitem [{\citenamefont {Hilfer}(2002)}]{hilfer2002}%
  \BibitemOpen
  \bibfield  {author} {\bibinfo {author} {\bibfnamefont {R.}~\bibnamefont
  {Hilfer}},\ }\bibfield  {title} {\bibinfo {title} {\textit{{H}} -function
  representations for stretched exponential relaxation and non-{Debye}
  susceptibilities in glassy systems},\ }\href
  {https://doi.org/10.1103/PhysRevE.65.061510} {\bibfield  {journal} {\bibinfo
  {journal} {Physical Review E}\ }\textbf {\bibinfo {volume} {65}},\ \bibinfo
  {pages} {061510} (\bibinfo {year} {2002})}\BibitemShut {NoStop}%
\bibitem [{\citenamefont {Mathai}\ \emph {et~al.}(2010)\citenamefont {Mathai},
  \citenamefont {Saxena},\ and\ \citenamefont {Haubold}}]{mathai2010}%
  \BibitemOpen
  \bibfield  {author} {\bibinfo {author} {\bibfnamefont {A.}~\bibnamefont
  {Mathai}}, \bibinfo {author} {\bibfnamefont {R.~K.}\ \bibnamefont {Saxena}},\
  and\ \bibinfo {author} {\bibfnamefont {H.~J.}\ \bibnamefont {Haubold}},\
  }\href {https://doi.org/10.1007/978-1-4419-0916-9} {\emph {\bibinfo {title}
  {The {H}-{Function}}}}\ (\bibinfo  {publisher} {Springer},\ \bibinfo
  {address} {New York, NY},\ \bibinfo {year} {2010})\BibitemShut {NoStop}%
\bibitem [{\citenamefont {Lukyanov}\ and\ \citenamefont
  {Zamolodchikov}(2001)}]{lukyanov_soliton_ff}%
  \BibitemOpen
  \bibfield  {author} {\bibinfo {author} {\bibfnamefont {S.}~\bibnamefont
  {Lukyanov}}\ and\ \bibinfo {author} {\bibfnamefont {A.~B.}\ \bibnamefont
  {Zamolodchikov}},\ }\bibfield  {title} {\bibinfo {title} {Form factors of
  soliton creating operators in the sine-gordon model},\ }\href@noop {}
  {\bibfield  {journal} {\bibinfo  {journal} {Nuclear Physics B}\ }\textbf
  {\bibinfo {volume} {607}},\ \bibinfo {pages} {437} (\bibinfo {year}
  {2001})}\BibitemShut {NoStop}%
\bibitem [{\citenamefont {Takamoto}\ \emph {et~al.}(2010)\citenamefont
  {Takamoto}, \citenamefont {Muraoka},\ and\ \citenamefont
  {Idogaki}}]{takamoto_2010}%
  \BibitemOpen
  \bibfield  {author} {\bibinfo {author} {\bibfnamefont {M.}~\bibnamefont
  {Takamoto}}, \bibinfo {author} {\bibfnamefont {Y.}~\bibnamefont {Muraoka}},\
  and\ \bibinfo {author} {\bibfnamefont {T.}~\bibnamefont {Idogaki}},\
  }\bibfield  {title} {\bibinfo {title} {The critical behaviour of the
  1-dimensional {XY} model with power law decay long range interactions},\
  }\href {https://doi.org/10.1088/1742-6596/200/2/022064} {\bibfield  {journal}
  {\bibinfo  {journal} {Journal of Physics: Conference Series}\ }\textbf
  {\bibinfo {volume} {200}},\ \bibinfo {pages} {022064} (\bibinfo {year}
  {2010})}\BibitemShut {NoStop}%
\end{thebibliography}
%

\end{document}